\documentclass[journal]{IEEEtran}
\usepackage{amsmath, amsthm, amsfonts}
\usepackage{bm}
\usepackage{graphicx,float,wrapfig}
\usepackage{enumerate}
\usepackage{cite,citesort}
\newtheorem{thm}{Theorem}[section]

\newtheorem{lem}[thm]{Lemma}

\theoremstyle{definition}
\newtheorem{defn}[thm]{Definition}
\theoremstyle{remark}

\newcommand{\cqfd}{\hfill\rule{2mm}{2mm}\medbreak\indent}

\DeclareMathOperator*{\esssup}{ess\,sup}
\DeclareMathOperator*{\essinf}{ess\,inf}
\DeclareMathOperator*{\ess}{ess}
\DeclareMathOperator*{\argmax}{arg\,max}
\DeclareMathOperator*{\argmin}{arg\,min}

\def\R{\mathbb{R}}
\def\Z{\mathbb{Z}}

\def \N {\mathbb{N}}

\title{On the Asymptotic Equivalence \\ of Circulant and Toeplitz Matrices}
\author{Zhihui Zhu and
        Michael B. Wakin
\thanks{
This work was supported by NSF grant CCF-1409261.}
\thanks{Z. Zhu and M. B. Wakin are with the Department
of Electrical Engineering and Computer Science, Colorado School of Mines, Golden, CO 80401 USA. e-mail: \{zzhu,mwakin\}@mines.edu.}
}

\begin{document}

\maketitle
\begin{abstract}
Any sequence of uniformly bounded $N\times N$ Hermitian Toeplitz matrices $\{\bm H_N\}$ is asymptotically equivalent to a certain sequence of $N\times N$ circulant matrices $\{\bm C_N\}$ derived from the Toeplitz matrices in the sense that $\left\|\bm H_N - \bm C_N \right\|_F = o(\sqrt{N})$ as $N\rightarrow \infty$. This implies that certain collective behaviors of the eigenvalues of each Toeplitz matrix are reflected in those of the corresponding circulant matrix and supports the utilization of the computationally efficient fast Fourier transform (instead of the Karhunen-Lo\`{e}ve transform) in applications like coding and filtering. In this paper, we study the asymptotic performance of the {\it individual eigenvalue estimates}. We show that the asymptotic equivalence of the circulant and Toeplitz matrices implies the individual asymptotic convergence of the eigenvalues for certain types of Toeplitz matrices. We also show that these estimates asymptotically approximate the largest and smallest eigenvalues for more general classes of Toeplitz matrices. 
\end{abstract}


\section{Introduction}\label{SzegoThm}

\subsection{Szeg\H{o}'s Theorem}

Toeplitz matrices are of considerable interest in statistical signal processing and information theory~\cite{grenander1958toeplitz,gray1972asymptotic,pearl1973coding,makhoul1975linear,kailath1978inverses}. An $N\times N$ Toeplitz matrix $\bm{H}_N$ has the form\footnote{Through the paper, finite-dimensional vectors and matrices are indicated by bold characters and we index such vectors and matrices beginning at $0$.}
\begin{align*}
\bm{H}_N=\left[\begin{array}{ccccc}h[0] & h[-1] & h[-2] & \ldots & h[-(N-1)] \\ h[1] & h[0] & h[-1] & & \\ h[2] & h[1] & h[0] & & \vdots \\ \vdots & & & \ddots & \\ h[N-1] & & \cdots & & h[0]\end{array}\right]
\end{align*}
or $\bm{H}_N[m,n] = h[m-n];m,n \in [N]:= \{0,1,\ldots,N-1\}$. The covariance matrix of a random vector obtained by sampling a wide-sense stationary (WSS) random process is an example of such a matrix.

Throughout this paper, we consider $\bm H_N$ that is Hermitian, i.e., $\bm H_N^H = \bm H_N$, and we suppose that the eigenvalues of $\bm H_N$ are denoted and arranged as $\lambda_0(\bm H_N)\geq \cdots\geq \lambda_{N-1}(\bm H_N)$. Here the Hermitian transpose of a matrix $\bm A$ is denoted by $\bm A^H$.

Szeg\H{o}'s theorem~\cite{grenander1958toeplitz} describes the collective asymptotic behavior (as $N\rightarrow \infty$) 
of the eigenvalues of a sequence of Hermitian Toeplitz matrices $\{\bm{H}_N\}$ by defining a function $\widetilde{h}(f)\in L^2([0,1])$ with Fourier series\footnote{This can also be interpreted using the discrete-time Fourier transform (DTFT). That is, we can define $\widehat h(f) = \sum_{k=-\infty}^{\infty}h[k]e^{-j2\pi kf} = \widetilde h(1-f)$. However, it is more common to view $h$ as the Fourier series of the symbol $\widetilde{h}$; see~\cite{grenander1958toeplitz,deift2012Spectrum}.}
\begin{equation}
\begin{split}
h[k] =& \int_{0}^{1}\widetilde{h}(f)e^{-j2\pi kf}df,~~k\in\Z,\\
\widetilde{h}(f) =& \sum_{k=-\infty}^{\infty}h[k]e^{j2\pi kf},~~f\in[0,1].
\end{split}
\nonumber\end{equation}
Usually $\widetilde{h}(f)$ is referred to as the symbol or generating function for the $N\times N$ Toeplitz matrices $\{\bm{H}_N\}$.

Suppose $\widetilde{h}\in L^{\infty}([0,1])$. Szeg\H{o}'s theorem~\cite{grenander1958toeplitz} states that
\begin{equation}
\lim_{N\rightarrow \infty}\frac{1}{N}\sum_{l=0}^{N-1}\vartheta(\lambda_l(\bm{H}_N)) = \int_0^{1}\vartheta(\widetilde{h}(f))df,
\label{eq:Szego thm}\end{equation}
where $\vartheta$ is any function continuous on the range of $\widetilde h$. As one example, choosing $\vartheta (x) = x$ yields
\begin{align*}
\lim_{N\rightarrow \infty}\frac{1}{N}\sum_{l=0}^{N-1}\lambda_l(\bm{H}_N) = \int_0^{1}\widetilde h(f)df.
\end{align*}
In words, this says that as $N\rightarrow \infty$, the average eigenvalue of $\bm{H}_N$ converges to the average value of the symbol $\widetilde{h}(f)$ that generates $\bm{H}_N$. As a second example, suppose $\widetilde h(f)>0$ (and thus $\lambda_l(\bm H_N)>0$ for all $l\in[N]$ and $N \in \N$) and let $\vartheta$ be the $\log$ function. Then Szeg\H{o}'s theorem indicates that
\begin{align*}
\lim_{N\rightarrow \infty}\frac{1}{N}\log\left(\det\left( \bm H_N\right) \right) = \int_0^{1}\log\left(\widetilde h(f)\right) df.
\end{align*}
This relates the determinant of the Toeplitz matrix to its symbol.

Szeg\H{o}'s theorem has been widely used in the areas of signal processing, communications, and information theory. A paper and review by Gray~\cite{gray1972asymptotic,gray2005toeplitz} serve as a remarkable elementary introduction in the engineering literature and offer a simplified proof of Szeg\H{o}'s original theorem. The result has also been extended in several ways. For example, the Avram-Parter theorem~\cite{avram1988bilinear,parter1986distribution}, a generalization of Szeg\H{o}'s theorem, relates the collective asymptotic behavior of the singular values of a general (non-Hermitian) Toeplitz matrix to the absolute value of its symbol, i.e., $|\widetilde h(f)|$. Tyrtyshnikov~\cite{tyrtyshnikov1996unifying} proved that Szeg\H{o}'s theorem holds if $\widetilde h(f)\in \R$ and $\widetilde h(f)\in L^{2}([0,1])$, and Zamarashkin and Tyrtyshnikov~\cite{zamarashkin1997distribution} further extended Szeg\H{o}'s theorem to the case when $\widetilde h(f)\in \R$ and $\widetilde h(f)\in L^{1}([0,1])$. Sakrison~\cite{sakrison1969extension} extended Szeg\H{o}'s theorem to high dimensions. Gazzah et al.~\cite{gazzah2001asymptotic} and Guti\'{e}rrez-Guti\'{e}rrez and Crespo~\cite{gutierrez2008asympBolckToeplitz} extended Gray's results on Toeplitz and circulant matrices to block Toeplitz and block circulant matrices and derived Szeg\H{o}'s theorem for block Toeplitz matrices.

Most relevant to our work, Bogoya et al.~\cite{Bogoya2015maximum} studied the {\em individual} asymptotic behavior of the eigenvalues of Toeplitz matrices by interpreting Szeg\H{o}'s theorem in probabilistic language. In the case that the range of $\widetilde h$ is connected, Bogoya et al.\ related the eigenvalues to the values obtained by sampling the symbol $\widetilde{h}(f)$ uniformly in frequency on $[0,1]$.

\subsection{Motivation}

Despite the power of Szeg\H{o}'s theorem, in many scenarios (such as certain coding and filtering applications~\cite{gray1972asymptotic,pearl1973coding}), one may only have access to $\bm{H}_N$ and not $\widetilde{h}$. In such cases, it is still desirable to have practical and efficiently computable estimates of individual eigenvalues of $\bm{H}_N$. We elaborate on two example applications below.

\vspace{0.2cm}
{\bf $i.$ Estimating the condition number of a positive-definite Toeplitz matrix.} The linear system $\bm H_N \bm y = \bm b$ arises naturally in many signal processing and estimation problems such as linear prediction~\cite{makhoul1975linear,kailath1978inverses}. The condition number $\kappa(\bm H_N)$ of the Toeplitz matrix $\bm H_N$ is important when solving such systems. For example, the speed of solving such linear systems via the widely used conjugate gradient method is determined by the condition number: the larger $\kappa(\bm H_N)$, the slower the convergence of the algorithm. In case of large $\kappa(\bm H_N)$, preconditioning can be applied to ensure fast convergence. Thus estimating the smallest and largest eigenvalues of a symmetric positive-definite Toeplitz matrix (such as the covariance matrix of a random vector obtained by sampling a stationary random process) is of considerable interest~\cite{dembo1988bounds,laudadio2008lowerBoundToeplitz}.

\vspace{0.2cm}
{\bf $ii.$  Spectrum sensing algorithm for cognitive radio.}
Spectrum sensing is a fundamental task in cognitive ratio, which aims to best utilize the available spectrum by identifying unoccupied bands~\cite{mitola1999cognitive,haykin2005cognitive,zeng2009eigenvalueSpectrumSensing}. Zeng and Ling~\cite{zeng2009eigenvalueSpectrumSensing} have proposed spectrum sensing methods for cognitive radio based on the eigenvalues of a Toeplitz covariance matrix. These eigenvalue-based algorithms overcome the noise uncertainty problem which exists in alternative methods based on energy detection.

\vspace{0.2cm}
Aside from the above applications, approximate and efficiently computable eigenvalue estimates can also be used as the starting point for numerical algorithms that iteratively compute eigenvalues with high precision.

\subsection{Contributions}

In this paper, we consider estimates for the eigenvalues of a Toeplitz matrix that are obtained through a two-step process:
\begin{enumerate}
\item Transform the Toeplitz matrix into a circulant matrix using a certain procedure described below.
\item Compute the eigenvalues of the circulant matrix.
\end{enumerate}
Both of these steps can be performed efficiently; in particular, the eigenvalues of an $N \times N$ circulant matrix can be computed in $O(N\log N)$ time\footnote{We say $g_1 (N) = O(g_2(N))$ if and only if there exist a positive real number $t$ and $M\in \N$ such that $g_1(N)\leq t g_2(N)$ for all $N\geq M$.} using the fast Fourier transform (FFT). The individual eigenvalues of the circulant matrix approximate those of the Toeplitz matrix. We study the quality of this approximation.

An $N\times N$ circulant matrix $\bm C_N$ is a special Toeplitz matrix of the form
\begin{align*} \bm{C}_N=\left[\begin{array}{ccccc}c[0] & c[1] & c[2] & \ldots & c[N-1] \\ c[N-1] & c[0] & c[1] & & \\ c[N-2] & c[N-1] & c[0] & & \vdots \\ \vdots & & & \ddots & \\ c[1] & & \cdots & & c[0]\end{array}\right].
\end{align*}
Circulant matrices arise naturally in applications involving the discrete Fourier transform (DFT)~\cite{pearl1973coding}; in particular, any circulant matrix can be unitarily diagonalized using the DFT matrix. Circulant matrices offer a nontrivial but simple set of objects that can be used for problems involving Toeplitz matrices. For example, the product $\bm H_N\bm x$ can be computed in $O(N\log N)$ time by embedding $\bm H_N$ into a $(2N-1)\times (2N-1)$ circulant matrix and using the FFT to perform matrix-vector multiplication. Also Gray~\cite{gray1972asymptotic,gray2005toeplitz} showed that Toeplitz and circulant matrices are asymptotically equivalent in a certain sense; this implies that their eigenvalues have similar {\it collective behavior}. See Section~\ref{sec:preliminaries} for formal definitions. Finally, we note that circulant matrices have been used as preconditioners~\cite{strang1986proposal,chan1988optimal} of Toeplitz matrices in iterative methods for solving linear systems of the form $\bm H_N \bm y = \bm b$.

We consider estimates for the eigenvalues of a Toeplitz matrix obtained from a well-constructed circulant matrix. The eigenvalues of the circulant matrix can be computed efficiently without constructing the whole matrix; one merely applies the FFT to the first row of the matrix. We do {\em not} provide new circulant approximations to Toeplitz matrices in this paper; rather we sharpen the analysis on the asymptotic equivalence of Toeplitz and certain circulant matrices~\cite{gray1972asymptotic,pearl1973coding,gray2005toeplitz} by establishing results in terms of {\it individual eigenvalues} rather than collective behavior. To the best of our knowledge, this is the first work that provides guarantees for asymptotic equivalence in terms of individual eigenvalues.

\subsection{Circulant Approximations to $\bm H_N$}
\label{subsec:circapproxdefs}

We consider the following circulant approximations that have been widely used in information theory and applied mathematics.

\subsubsection{$\widetilde{\bm{C}}_{N}$}

Bogoya et al.~\cite{Bogoya2015maximum} proved that the samples of the symbol $\widetilde h$ are the main asymptotic terms of the eigenvalues of the Toeplitz matrix $\bm H_N$. Given only $\bm H_N$, one practical strategy for estimating the eigenvalues is to first approximate $\widetilde{h}$ by the $\left(N-1\right)^{\text{th}}$ partial Fourier sum $S_{N-1}(f) = \sum_{k=-(N-1)}^{N-1}h[k]e^{j2\pi f k}$. Then construct a circulant matrix whose eigenvalues are samples of $S_{N-1}(f)$, i.e., $S_{N-1}(\frac{ l}{N})$. We let $\widetilde{\bm{C}}_{N}$ denote the corresponding circulant matrix, whose top row $\left(\widetilde c[0], \widetilde c[1], \ldots, \widetilde c[N-1] \right)$ can be obtained as
    \begin{align*}
    \widetilde c[k] &= \frac{1}{N}\sum_{n=0}^{N-1}S_{N-1}(\frac{n}{N})e^{j2\pi kn/N}\\
    & = \frac{1}{N}\sum_{n=0}^{N-1}\sum_{k'=-(N-1)}^{N-1}h[k']e^{j2\pi (k+k')n/N}\\
    & = \sum_{k'=-(N-1)}^{N-1}h[k']\left(\sum_{n=0}^{N-1}\frac{1}{N}e^{j2\pi (k+k')n/N}\right)\\
    & = \left\{\begin{array}{ll}h[0],& k = 0,\\h[-k]+h[N-k], & k = 1,2,\ldots,N-1,\end{array}\right.
    \end{align*}
where the last line utilizes the fact
$$
\sum_{n=0}^{N-1}\frac{1}{N}e^{j2\pi (k+k')n/N} = \left\{\begin{array}{ll}1, & \mbox{mod}(k+k',N) = 0,\\0, & \mbox{otherwise}.\end{array}  \right.
$$

\subsubsection{$\widehat {\bm C}_N$}

Following the same strategy, we first compute the $\left( \left\lfloor\frac{N-1}{2}\right\rfloor\right)^{\text{th}}$ partial Fourier sum $$S_{\lfloor\frac{N-1}{2}\rfloor}(f) = \sum_{k=-\left\lfloor\frac{N-1}{2}\right\rfloor}^{\left\lfloor\frac{N-1}{2}\right\rfloor}h[k]e^{j2\pi f k}.$$
Let $\widehat {\bm C}_N$ denote the $N\times N$ circulant matrix whose eigenvalues are samples of $S_{\left\lfloor\frac{N-1}{2}\right\rfloor}(f) $, i.e, $S_{\left\lfloor\frac{N-1}{2}\right\rfloor}(\frac{ l}{N}) $. With simple manipulations, the top row $\left(\widehat c[0], \widehat c[1], \ldots, \widehat c[N-1] \right)$ of $\widehat{\bm C}_N$ is given by
\begin{align*}
\widehat c[k] = \left\{\begin{array}{ll} h[-k], & 0\leq k\leq  \lfloor \frac{N-1}{2}\rfloor, \\h[N-k], &  \lceil \frac{N+1}{2}\rceil \leq k<N,\\
0, & k = N/2, \end{array}\right.
\end{align*}
when $N$ is even, and
\begin{align*}
\widehat c[k] = \left\{\begin{array}{ll} h[-k], & 0\leq k\leq \lfloor \frac{N-1}{2}\rfloor,\\h[N-k], & \lceil \frac{N+1}{2}\rceil\leq k<N, \end{array}\right.
\end{align*}
when $N$ is odd.

Strang~\cite{strang1986proposal} first employed such circulant matrices as preconditioners to speed up the convergence of iterative methods for solving Toeplitz linear systems. This approach is quite simple. The underlying idea is that the sequence $h[k]$ usually decays quickly as $k$ grows large, and thus we keep the largest part of the Toeplitz matrix and fill in the remaining part to form a circulant approximation.

\subsubsection{$\overline{\bm C}_N$}

In the Fourier analysis literature, it is known that Ces\`{a}ro sum has rather better convergence than the partial Fourier sum~\cite{korner1989fourier}. The $N^{\text{th}}$ Ces\`{a}ro sum is defined as
    $$\sigma_N(f) = \frac{\sum_{n=0}^{N-1}S_n(f)}{N}.$$
We use $\overline {\bm C}_N$ to denote the $N\times N$ circulant matrix whose eigenvalues are samples of $\sigma_N(f)$, i.e., $\sigma_N(\frac{l}{N})$. The top row ($\overline c[0], \overline c[1], \ldots, \overline c[N-1]$) of $\overline{\bm C}_N$ can be obtained as follows
\begin{align*}
\overline c[k] &= \frac{1}{N}\sum_{l=0}^{N-1}\sigma_N(\frac{l}{N})e^{j2\pi kl/N}\\
& = \frac{1}{N}\sum_{l=0}^{N-1}\frac{1}{N}\sum_{n=0}^{N-1}\sum_{k'=-n}^n h[k']e^{j2\pi l(k+k')/N}\\
& = \frac{1}{N}\sum_{n=0}^{N-1}\sum_{k'=-n}^n \left(h[k'] \sum_{l=0}^{N-1}\frac{1}{N}e^{j2\pi l(k+k')/N}\right)\\
& = \frac{1}{N}\left((N-k)h[-k] + k h[N-k]\right).
\end{align*}

Pearl~\cite{pearl1973coding} first analyzed such a circulant approximation and its applications in coding and filtering. The same circulant approximation (referred to as an optimal preconditioner) was also proposed by Chan~\cite{chan1988optimal}. The optimal preconditioner is the solution to the following optimization problem
\begin{align*}
\text{minimize}~ \|\bm C_N-\bm H_N\|_F
\end{align*}
over all $N\times N$ circulant matrices. One can verify that $\overline {\bm C}_N$ is the solution to the above problem.

\subsection{Main Results}
As a reminder, we assume throughout this paper that each $\bm H_N$ is Hermitian; this ensures that all $\bm C_N\in\left\{\widetilde {\bm C}_N, \widehat {\bm C}_N, \overline {\bm C}_N \right\}$ are Hermitian as well.
Let $\left\{\lambda_l\left(\bm C_N\right)\right\}_{l\in[N]}$ denote the eigenvalues of the circulant matrix $\bm C_N$ for all $\bm C_N\in\left\{\widetilde {\bm C}_N, \widehat {\bm C}_N, \overline {\bm C}_N \right\}$. Let $\lambda_l({\bm C}_{N})$  be permuted that such that $\lambda_{\rho(0)}({\bm C}_{N}) \geq \lambda_{\rho(1)}({\bm C}_{N}) \geq\cdots \geq \lambda_{\rho(N-1)}({\bm C}_{N})$. In this paper, we establish the following results.

\begin{thm}
Suppose that the sequence $h[k]$ is absolutely summable. Then
\begin{align}
\lim_{N\rightarrow\infty}\max_{l\in[N]}\left|\lambda_l(\bm{H}_N) - \lambda_{\rho(l)}({\bm C}_{N})\right| = 0,
\label{eq:individual equiv for continuous function}
\end{align}
for all $\bm C_N\in\left\{\widetilde {\bm C}_N, \widehat {\bm C}_N, \overline {\bm C}_N \right\}$.
\label{thm:individual eig toeplitz for continuous function}
\end{thm}

Theorem~\ref{thm:individual eig toeplitz for continuous function} states that the individual asymptotic convergence of the eigenvalues between the Toeplitz matrices $\bm H_N$ and circulant matrices $\bm C_N\in\left\{\widetilde {\bm C}_N, \widehat {\bm C}_N, \overline {\bm C}_N \right\}$ holds as long as $h[k]$ is absolutely summable. Its proof involves the uniform convergence of a Fourier series and the fact that the equal distribution of two sequences implies individual asymptotic equivalence of two sequences in a certain sense. By utilizing the Sturmian separation theorem~\cite{Horn1985MatrixAnalysis}, we also provide the convergence rate for band Toeplitz matrices as follows.
\begin{thm}
Suppose that $h[k] = 0$ for all $|k|> r$, i.e., $\bm H_N$ is a band Toeplitz matrix when $N > r$. Then
\begin{align}
\max_{l\in[N]}\left|\lambda_l(\bm{H}_N) - \lambda_{\rho(l)}({\bm C}_{N})\right| = O(\frac{1}{N})
\label{eq:convergence rate band Toeplitz}\end{align}
as $N\rightarrow \infty$ for all $\bm C_N\in\left\{\widetilde {\bm C}_N, \widehat {\bm C}_N, \overline {\bm C}_N \right\}$.
\label{thm:convergence rate band Toeplitz}\end{thm}

Utilizing the fact that the Ces\`{a}ro sum has rather better convergence than the partial Fourier sum, the following result establishes a weaker condition on $h[k]$ for the individual asymptotic convergence of the eigenvalues between $\bm H_N$ and $\overline{\bm C}_N$.

\begin{thm}
Suppose that $h[k]$ is square summable and $\widetilde{h}\in L^{\infty}([0,1])$ is Riemann integrable and the essential range of $\widetilde h$ is $\left[\essinf \widetilde h,\esssup\widetilde h \right]$, i.e., the essential range of $\widetilde h$ is connected. Then
\begin{align}
\lim_{N\rightarrow\infty}\max_{l\in[N]}\left|\lambda_l(\bm{H}_N) - \lambda_{\rho(l)}(\overline{\bm C}_{N})\right| = 0.
\label{eq: individual equiv}
\end{align}
\label{thm:individual eig toeplitz}
\end{thm}

Note that the sequence $h[k]$ being absolutely summable implies that $h[k]$ is square summable, that $\widetilde h\in L^{\infty}([0,1])$ is Riemann integrable, and that its range is connected~\cite{korner1989fourier} (actually $\widetilde h$ is continuous). However, the converse of this statement does not hold. We provide an example in Section~\ref{sec:example h[k] not summalbe but range is connected}.

Finally, the following result concerns the convergence of the largest and smallest eigenvalues for more general classes of Toeplitz matrices.
\begin{thm}
Suppose that $\widetilde{h}\in L^{\infty}([0,1])$ is Riemann integrable. Then
\begin{align*}
&\lim_{N\rightarrow \infty} \lambda_0\left(\bm H_N \right) = \lim_{N\rightarrow \infty} \lambda_{\rho(0)}\left( \overline{\bm C}_N \right) = \esssup\widetilde h, \\
&\lim_{N\rightarrow \infty} \lambda_{N-1}\left(\bm H_N \right) =  \lim_{N\rightarrow \infty} \lambda_{\rho(N-1)}\left( \overline{\bm C}_N \right) = \essinf\widetilde h.
\end{align*}
\label{thm:bounds extreme eigenvalues}
\end{thm}

\noindent{\bf Remark.} Theorem~\ref{thm:bounds extreme eigenvalues} works only for the circulant matrix $\overline{\bm C}_N$ and not $\widetilde{\bm C}_N$ or $\widehat{\bm C}_N$. This is closely related to the fact that the partial Ces\`{a}ro sum has better convergence than the partial Fourier sum~\cite{korner1989fourier}.

\noindent{\bf Remark.} Theorem \ref{thm:bounds extreme eigenvalues} only requires $\widetilde h$ to be bounded and Riemann integrable, while Theorem \ref{thm:individual eig toeplitz} requires the range of $\widetilde h$ to be connected.

Directly computing the eigenvalues of the Toeplitz matrix $\bm H_N$ would generally require $O(N^3)$ flops. By exploiting the special structure of Toeplitz matrices, Trench~\cite{trench1989numerical} presented an iterative algorithm (which combines the Levinson-Durbin algorithm with an iterative root-finding procedure) requiring $O(N^2)$ operations per eigenvalue. Laudadio et al.~\cite{laudadio2008lowerBoundToeplitz} summarized several algorithms to estimate the smallest eigenvalue of a symmetric positive-definite Toeplitz matrix. These algorithms need $O(N^2)$ flops. Luk and Qiao~\cite{luk2000fast} proposed a fast algorithm (which consists of a Lanczos-type tridiagonalization procedure and a QR-type diagonalization method) that computes the eigenvalues in $O(N^2\log N)$ operations. In contrast, it is clear from Section~\ref{subsec:circapproxdefs} that the top row of  $\bm C_N\in\left\{\widetilde {\bm C}_N, \widehat {\bm C}_N, \overline {\bm C}_N \right\}$ can be computed at most in $O(N)$ flops because of the closed-form expressions. Computing the eigenvalues for the corresponding $N\times N$ circulant matrices via the FFT requires only $O(N\log N)$ flops, and at the same time, Theorems~\ref{thm:individual eig toeplitz for continuous function}--\ref{thm:bounds extreme eigenvalues} ensure that these eigenvalues are asymptotically equivalent to the eigenvalues of the Toeplitz matrix.

The above results---characterizing the individual asymptotic convergence of the eigenvalues between Toeplitz and circulant matrices---serve as complements to the literature on asymptotic equivalence that has focused on the collective behavior of the eigenvalues. Before moving on, we briefly review said literature. In~\cite{gray1972asymptotic,gray2005toeplitz}, Gray showed the asymptotic equivalence\footnote{We define asymptotically equivalent sequences of matrices in Section~\ref{sec:preliminaries}.} of $\{\bm H_N\}$ and $\{\widetilde{\bm C}_N\}$ when the sequence $h[k]$ is absolutely summable. Pearl showed the asymptotic equivalence of $\{\bm H_N\}$ and $\{\overline{\bm C}_N\}$ when the sequence $h[k]$ is square summable and $\bm H_N$ and $\overline{\bm C}_N$ have bounded eigenvalues for all $N\in\N$. The spectrum of the preconditioned matrix $\bm C_N^{-1}\bm H_N$ asymptotically clustering around one was investigated in~\cite{chan1989circulant,chan1989toeplitz,chan1991spectra,tyrtyshnikov1996unifying}.

Finally, as noted previously, Bogoya et al.~\cite{Bogoya2015maximum} studied the {\em individual} asymptotic behavior of the eigenvalues of Toeplitz matrices by interpreting Szeg\H{o}'s theorem in probabilistic language. To the best of our knowledge, \cite{Bogoya2015maximum} was the first work that provided conditions under which Szeg\H{o}'s theorem implies individual asymptotic eigenvalue estimates by sampling the symbol $\widetilde{h}(f)$ uniformly in frequency on $[0,1]$.  Our estimates for the eigenvalues of a Toeplitz matrix differ from~\cite{Bogoya2015maximum} in that they are only dependent on the entries of $\bm H_N$ (instead of the symbol $\widetilde{h}(f)$).  As we attempt to forge a connection between equal distribution of two sequences of matrices and individual asymptotic equivalence of the eigenvalues, we bridge the gap between equal distribution of two real sequences (see Definition~\ref{def:Weyl equal distribution}) and individual asymptotic equivalence of the two real sequences (see Definition~\ref{def:individual convergence}). In particular, we provide two conditions under which equal distribution implies individual asymptotic equivalence in Theorems~\ref{thm:uniform to individual for contunuous fuc} and \ref{thm:uniform to individual}, which may themselves be of independent interest. For our proof, as motivated by~\cite{Bogoya2015maximum},  we utilize the same approach of interpreting asymptotic equivalence in probabilistic language. However, \cite{Bogoya2015maximum}~involves the quantile function, while our approach involves only the cumulative distributive function and works on the sequences directly with proof by contradiction. Moreover,~\cite{Bogoya2015maximum} requires the sequences of the eigenvalues to be strictly inside the range of $\widetilde{h}$,\footnote{It is unclear whether Szeg\H{o}'s theorem implies individual asymptotic eigenvalue estimates by sampling the symbol $\widetilde{h}(f)$ uniformly if the eigenvalues are outside the range of $\widetilde{h}$. For this case, besides a connected range for $\widetilde h$, we suspect more conditions are needed to ensure the results in~\cite{Bogoya2015maximum} still hold.} while our work covers more general cases where the sequences of the eigenvalues can be outside of the range of $\widetilde{h}$ (since the eigenvalues of $\widehat {\bm C}_N$ and $\widetilde {\bm C}_N$ can be outside the range of $\widetilde h$) as illustrated in Theorem~\ref{thm:uniform to individual for contunuous fuc}. See also our remark at the end of Section~\ref{sec:proofremark}.
Finally, while Theorem~\ref{thm:uniform to individual} in the present work requires two sequences to be inside the essential range of a function, the range of this function is not required to be connected; connectedness is needed in~\cite{Bogoya2015maximum} so that the quantile function defined there is uniformly continuous.

The rest of the paper is organized as follows. Section~\ref{sec:preliminaries} states preliminary results on the asymptotic equivalence of Toeplitz and circulant matrices. We prove our main results in Section~\ref{sec:proof of theorems}. Section~\ref{sec:simulations} presents examples to illustrate our results, and Section~\ref{sec:conclusions} concludes the paper.

\section{Preliminaries}\label{sec:preliminaries}

Before proceeding, we introduce some notation used throughout the paper. Let $\mathcal{R}\left(\cdot \right)$ be the range of a function and  $\ess\mathcal{R}\left(\cdot \right)$ be the essential range of a function. Suppose $g(x):\R\rightarrow \R$. Then
\[
\ess\mathcal{R}(g) = \left\{y\in \R\mid \forall \epsilon>0:\mu(\left\{x:|g(x)-y|<\epsilon\right\})>0\right\},
\]
where $\mu(\cdot)$ is the Lebesgue measure of a set (i.e., the length of an interval when the set is an interval). For any $\Omega\subset\R$, let $\text{int}\left( \Omega\right)$ be the interior of the set $\Omega$. We say the essential range of the function $g(x):\R\rightarrow \R$ is connected if its essential range is a real interval (a set of real numbers that any number lies between two numbers in the set is also included in the set).

\subsection{Asymptotically Equivalent Matrices}

We begin with the notion of equal distribution of two real sequences, using a definition attributed to Weyl~\cite{grenander1958toeplitz}.

\begin{defn}[equal distribution \cite{grenander1958toeplitz}]
Assume that the sequences $\{ \{u_{N,l}\}_{l\in [N]} \}_{N=1}^{\infty}$ and  $\{ \{v_{N,l}\}_{l\in [N]} \}_{N=1}^{\infty}$ are absolutely bounded, i.e., there exist $a,b$ such that  $a \leq u_{N,l} \leq b$ and $a \leq v_{N,l}\leq b$ for all $l\in[N]$ and $N\in \N$. Then  $\{ \{u_{N,l}\}_{l\in [N]} \}_{N=1}^{\infty}$ and  $\{ \{v_{N,l}\}_{l\in [N]} \}_{N=1}^{\infty}$ are {\it equally distributed} if
\begin{align*}
\lim_{N\rightarrow \infty}\frac{1}{N}\sum_{l=0}^{N-1}\left(\vartheta \left(u_{N,l}\right)  - \vartheta \left(v_{N,l}\right)  \right) = 0.
\end{align*}
for every continuous function $\vartheta$ on $[a,b]$.
\label{def:Weyl equal distribution}
\end{defn}
We define the notion of individual asymptotic equivalence of two real sequences as follows.
\begin{defn}[individual asymptotic equivalence]
Assume that the sequences $\{ \{u_{N,l}\}_{l\in [N]} \}_{N=1}^{\infty}$ and  $\{ \{v_{N,l}\}_{l\in [N]} \}_{N=1}^{\infty}$ are arranged in decreasing order and are absolutely bounded, i.e., there exist $a,b$ such that  $a \leq u_{N,N-1} \leq \cdots \leq u_{N,1} \leq u_{N,0} \leq b$ and $a \leq v_{N,N-1} \leq \cdots \leq v_{N,1} \leq v_{N,0} \leq b$ for all $N\in \N$. Then  $\{ \{u_{N,l}\}_{l\in [N]} \}_{N=1}^{\infty}$ and  $\{ \{v_{N,l}\}_{l\in [N]} \}_{N=1}^{\infty}$ are {\it individually asymptotically equivalent} if
\begin{align*}
\lim_{N\rightarrow \infty}\max_{l\in[N]}\left|u_{N,l}  - v_{N,l}\right|  = 0.
\end{align*}
\label{def:individual convergence}
\end{defn}
We note that individual asymptotic equivalence is stronger than equal distribution since if $\{ \{u_{N,l}\}_{l\in [N]} \}_{N=1}^{\infty}$ and  $\{ \{v_{N,l}\}_{l\in [N]} \}_{N=1}^{\infty}$ are individually asymptotically equivalent, they are equally distributed (see Appendix~\ref{prf:uniform to individual for contunuous fuc}, where it is proved that \eqref{eq:individual convergence for continuous func} implies \eqref{eq:uniform convergence for continuous func} without using any of the assumptions of Theorem~\ref{thm:uniform to individual for contunuous fuc}). However, equal distribution in general does not imply individual asymptotic convergence of two real sequences. As a simple example, let $u_{N,0} =1, u_{N,l} = 0, v_{N,0} = 2, v_{N,l} = 0$ for all $l\in\{1,2,\ldots,N-1\}$ and $N\in\N$. We have
\begin{align*}
&\lim_{N\rightarrow \infty}\frac{1}{N}\sum_{l=0}^{N-1}\left(\vartheta \left(u_{N,l}\right)  - \vartheta \left(v_{N,l}\right)  \right) \\&= \lim_{N\rightarrow \infty}\frac{1}{N} (\vartheta(1) - \vartheta(2)) = 0
\end{align*}
for every continuous function $\vartheta$ on $[0,2]$. However, these two sequences are not individually asymptotically convergent since
\begin{align*}
\max_{l\in[N]}\left|u_{N,l}  - v_{N,l}\right| = 1
\end{align*}
for any $N\in\N$.

The asymptotic equivalence of two sequences of matrices is defined as follows.
\begin{defn}\cite{gray1972asymptotic,gray2005toeplitz}
Two sequences of $N\times N$ matrices $\{\bm A_N\}$ and $\{\bm B_N \}$ (where $\bm A_N$ and $\bm B_N$ denote $N\times N$ matrices) are said to be {\it asymptotically equivalent} if
$$\lim_{N\rightarrow \infty}\frac{\left\| \bm A_N - \bm B_N\right\|_F}{\sqrt{N}} = 0$$
and $\{\bm A_N\}$ and $\{\bm B_N \}$ are uniformly absolutely bounded\footnote{We say a sequence of matrices $\{\bm A_N\}$ is uniformly absolutely bounded if there exists a constant $M$ such that $\|\bm A_N\|_2\leq M$ is for all $N\in\N$.}, i.e.,
there exists a constant $M<\infty$ such that
$$\left\|\bm A_N \right\|_2, \left\|\bm B_N \right\|_2\leq M, \quad \forall N\in \N.$$
\label{def:matrix equivalents}
\end{defn}

Following the convention in Gray's~monograph~\cite{gray2005toeplitz}, we write $\bm A_N \sim \bm B_N$ if $\{\bm A_N\}$ and $\{\bm B_N\}$ are asymptotically equivalent. This kind of asymptotic equivalence is transitive, i.e., if $\bm A_N\sim \bm B_N$ and $\bm B_N \sim \bm C_N$, then $\bm A_N \sim \bm C_N$. Additional properties of $\sim$ can be found in~\cite{gray2005toeplitz}. The following result concerns the asymptotic eigenvalue behavior of asymptotically equivalent Hermitian matrices.

\begin{thm}
\cite[Theorem 2.4]{gray2005toeplitz} Let $\{\bm A_N\}$ and $\{\bm B_N\}$ be asymptotically equivalent sequences of Hermitian matrices with eigenvalues $\{\left\{\lambda_l\left(\bm A_N \right)\right\}_{l\in[N]}\}_{N=1}^{\infty}$ and $\{\left\{\lambda_l\left(\bm B_N \right)\right\}_{l\in[N]}\}_{N=1}^{\infty}$. Then there exist constants $a$ and $b$ such that
$$ a\leq\lambda_l\left(\bm A_N \right), \lambda_l\left(\bm B_N \right)\leq b,\enskip \forall\enskip l\in[N], N\in\N.$$
Let $\vartheta$ be any function continuous on $[a,b]$. We have
\begin{align*}
\lim_{N\rightarrow \infty}\frac{1}{N}\sum_{l=0}^{N-1}\left(\vartheta \left(\lambda_l\left(\bm A_N \right)\right)  - \vartheta \left(\lambda_l\left(\bm B_N \right)\right)  \right) = 0.
\end{align*}
\label{thm:equivalence of matrices}
\end{thm}

In light of this theorem, Definition~\ref{def:matrix equivalents} can be viewed as the matrix equivalent of Definition~\ref{def:Weyl equal distribution}. One can also define the matrix equivalent of Definition~\ref{def:individual convergence}, although we will not need this.

\subsection{Asymptotic Equivalence of Circulant and Toeplitz Matrices}

Any circulant matrix $\bm C_N$ is characterized by its top row. Let
\[
\bm{e}_f:=\left[\begin{array}{c}e^{j2\pi f0}\\e^{j2\pi f1}\\\vdots\\e^{j2\pi f(N-1)} \end{array}\right] \in \mathbb{C}^N, ~ f\in[0,1]
\]
denote a length-$N$ vector of samples from a discrete-time complex exponential signal with digital frequency $f$. Note that
\begin{align*}
\left(\bm C_N \bm e_{(N-l)/N}\right)[k] = & \sum_{n=0}^{N-1}c[n]e^{j2\pi (N-l) \left(k+n\right)/N}\\ = &e^{j2\pi (N-l) k/N}\left(\sum_{n=0}^{N-1}c[n]e^{-j2\pi l n/N}\right),
\end{align*}
which implies that
$$\bm C_N \bm e_{(N-l)/N} = \left(\sum_{n=0}^{N-1}c[n]e^{-j2\pi l n/N}\right) \bm e_{(N-l)/N} .$$
Thus the normalized DFT basis vectors $\left\{ \frac{1}{\sqrt{N}}\bm e_{l/N}\right\}_{l\in [N]}$ are the eigenvectors of any circulant matrix $\bm C_N$, and the corresponding eigenvalues are obtained by taking the DFT of the first row of $\bm C_N$. Specifically,
$$
\lambda_l\left(\bm C_N \right) = \sum_{n=0}^{N-1}c[n]e^{-j2\pi l n/N},
$$
which can be computed efficiently via the FFT. We note that $\left\{\lambda_l\left(\bm C_N\right)\right\}_{l\in[N]}$ are not necessarily arranged in any particular order; namely, they do not necessarily decrease with $l$.

For a sequence of Toeplitz matrices $\{\bm H_N\}$ and their respective circulant approximations discussed in Section~\ref{subsec:circapproxdefs}, the following result establishes asymptotic equivalence in terms of the collective behaviors of the eigenvalues. As a reminder, we assume that each $\bm H_N$ is Hermitian; this ensures that all $\bm C_N\in\left\{\widetilde {\bm C}_N, \widehat {\bm C}_N, \overline {\bm C}_N \right\}$ are Hermitian as well.

\begin{lem}
Suppose that the sequence $h[k]$ is square summable and $\{\bm H_N\}, \{\widetilde {\bm C}_N\}, \{\widehat {\bm C}_N\}, \{\overline {\bm C}_N\}$ are uniformly absolutely bounded.
Then
\begin{align*}
\bm H_N\sim \widehat {\bm C}_N\sim\widetilde {\bm C}_N\sim\overline {\bm C}_N,
\end{align*}
and
\begin{align*}
\lim_{N\rightarrow \infty}\frac{1}{N}\sum_{l=0}^{N-1}\left( \vartheta(\lambda_l(\bm{H}_N)) -\vartheta(\lambda_l(\bm{C}_{N})) \right) = 0,
\end{align*}
where $\vartheta$ is any continuous function on $[a,b]$  and $\bm C_N\in\left\{\widetilde {\bm C}_N, \widehat {\bm C}_N, \overline {\bm C}_N \right\}$. Here $[a,b]$ is the smallest interval that covers all the eigenvalues of $\bm H_N, \widetilde {\bm C}_N, \widehat {\bm C}_N$, and $\overline {\bm C}_N$.
\label{lem:circulantApproxCollective}
\end{lem}

\noindent{\textbf{Proof.}} See Appendix~\ref{prf:circulantApproxCollective}.

A stronger result follows simply from the elementary view of Weyl's theory of equal distribution~\cite{trench2012elementary}, which is presented in Lemma~\ref{lem: absolute uniform converg}. As a reminder, we do assume that the eigenvalues of each Toeplitz matrix are ordered such that $\lambda_0(\bm H_N)\geq \cdots\geq \lambda_{N-1}(\bm H_N)$.

\begin{lem}  Suppose that  the sequence $h[k]$ is square summable and $\{\bm H_N\}, \{\widetilde {\bm C}_N\}, \{\widehat {\bm C}_N\}, \{\overline {\bm C}_N\}$ are uniformly absolutely bounded. Let $\lambda_l({\bm C}_{N})$  be permuted that such that $\lambda_{\rho(0)}({\bm C}_{N}) \geq \lambda_{\rho(1)}({\bm C}_{N}) \geq\cdots \geq \lambda_{\rho(N-1)}({\bm C}_{N})$. Then
$$
\lim_{N\rightarrow \infty}\frac{1}{N}\sum_{l=0}^{N-1}\left|\vartheta(\lambda_l(\bm{H}_N)) - \vartheta(\lambda_{\rho(l)}({\bm C}_{N}))\right| = 0
$$
for every  function $\vartheta$  that is continuous on $[a,b]$  and $\bm C_N\in\left\{\widetilde {\bm C}_N, \widehat {\bm C}_N, \overline {\bm C}_N \right\}$. Here $[a,b]$ is the smallest interval that covers all the eigenvalues of $\bm H_N, \widetilde {\bm C}_N, \widehat {\bm C}_N$, and $\overline {\bm C}_N$.
\end{lem}
\noindent{\textbf{Proof}. This result follows simply from Lemmas~\ref{lem:circulantApproxCollective} and~\ref{lem: absolute uniform converg}.  \cqfd

\section{Proofs of Main Theorems}\label{sec:proof of theorems}

\subsection{Proof of Theorem \ref{thm:individual eig toeplitz for continuous function}}

We first provide a strong condition under which the equal distribution of two sequences is equivalent to individual asymptotic equivalence.
\begin{thm}
Assume that the sequences  $\{ \{u_{N,l}\}_{l\in [N]} \}_{N=1}^{\infty}$ and  $\{ \{v_{N,l}\}_{l\in [N]} \}_{N=1}^{\infty}$ are absolutely bounded, i.e., there exist $a', b'$ such that  $b' \geq u_{N,0}\geq u_{N,1}\geq \cdots\geq u_{N,N-1} \geq a'$ and $b'\geq v_{N,0}\geq v_{N,1}\geq \cdots\geq v_{N,N-1}\geq a'$ for all $N\in \N$. Furthermore, suppose there exists a non-constant continuous function $g(x):[c,d]\rightarrow \mathbb R$ such that
\begin{align*}
\lim_{N\rightarrow \infty} u_{N,0} = \lim_{N\rightarrow \infty} v_{N,0} = \max_{x\in[c,d]}g(x),\\
\lim_{N\rightarrow \infty} u_{N,N-1} = \lim_{N\rightarrow \infty} v_{N,N-1} = \min_{x\in[c,d]}g(x),
\end{align*}
and
\begin{align*}
\lim_{N\rightarrow \infty}\frac{1}{N}\sum_{l=0}^{N-1}\vartheta(u_{N,l}) = \frac{1}{d-c}\int_{c}^d\vartheta(g(x))dx < \infty
\end{align*}
for every function $\vartheta$ that is continuous on $[a,b]$, where $[a,b]$ is the smallest interval that covers $[a',b']$ and the range of $g(x)$.  Then the following are equivalent:
\begin{align}
\lim_{N\rightarrow \infty}\frac{1}{N}\sum_{l=0}^{N-1}\left(\vartheta(u_{N,l}) - \vartheta(v_{N,l})\right) = 0;
\label{eq:uniform convergence for continuous func}\end{align}
\begin{align}
\lim_{N\rightarrow \infty} \max_{l \in [N]} \left| u_{N,l} - v_{N,l} \right| =0.
\label{eq:individual convergence for continuous func}\end{align}
\label{thm:uniform to individual for contunuous fuc}\end{thm}
\noindent{\textbf{Proof} (of Theorem \ref{thm:uniform to individual for contunuous fuc})}. See Appendix \ref{prf:uniform to individual for contunuous fuc}. \cqfd

If $\widetilde h(f) \equiv C$ is a constant function, then $\bm H_N$ and $\bm C_N$ are diagonal matrices with all diagonals being $C$  for all $\bm C_N\in\left\{\widetilde {\bm C}_N, \widehat {\bm C}_N, \overline {\bm C}_N \right\}$ and $N\in\mathbb N$. Thus
$\lambda_{l}\left(\bm H_N \right) = \lambda_{l}\left({\bm C}_N \right) = C$ for all $l\in [N]$  and $\bm C_N\in\left\{\widetilde {\bm C}_N, \widehat {\bm C}_N, \overline {\bm C}_N \right\}$. The following result establishes the range of the eigenvalues of $\overline {\bm C}_N$ and $\bm H_N$ for the case when $\widetilde h(f)$ is not a constant function.
\begin{lem}
Suppose that $\widetilde{h}\in L^{\infty}([0,1])$ and $\widetilde{h}$ is not a constant function. Also let $\lambda_l(\overline {\bm C}_N)$ be permuted such that $\lambda_{\rho(0)}(\overline {\bm C}_N) \geq \lambda_{\rho(1)}(\overline {\bm C}_N) \geq\cdots \geq \lambda_{\rho(N-1)}(\overline {\bm C}_N)$. Then
$$
\textup{ess}\inf \widetilde h < \lambda_{N-1}(\bm H_N) \leq \lambda_{\rho(N-1)}(\overline {\bm C}_N)
$$
and
$$
\lambda_{\rho(0)}(\overline {\bm C}_N)\leq\lambda_0(\bm H_N) < \textup{ess}\sup \widetilde h.
$$
\label{lem:bounds Pearl's eigenvalues}\end{lem}
\noindent{\textbf{Proof}} (of Lemma \ref{lem:bounds Pearl's eigenvalues}).
We first rewrite $\lambda_l\left(\overline {\bm C}_N \right)$ as
\begin{align*}
\lambda_l\left(\overline{\bm C}_N \right) &= \sum_{n=0}^{N-1}\overline c[n]e^{\frac{-j2\pi ln}{N}}\\
&=\sum_{n=0}^{N-1}\frac{1}{N}\left((N-n)h[-n] + n h[N-n]\right)e^{\frac{-j2\pi ln}{N}}\\
& = \left\langle \bm H_N\frac{1}{\sqrt{N}}\bm e_{l/N},  \frac{1}{\sqrt{N}}\bm e_{l/N}\right\rangle.
\end{align*}
By definition, $\lambda_0(\bm H_N) = \max_{\|\bm v\|_2 =1} \langle \bm H_N \bm v, \bm v\rangle$ and $\lambda_{N-1}(\bm H_N) = \min_{ \|\bm v\|_2 =1} \langle \bm H_N \bm v, \bm v\rangle$, we obtain
$$\lambda_{N-1}(\bm H_N) \leq \lambda_l(\overline {\bm C}_N)\leq \lambda_0(\bm H_N)  ,~\forall~l.$$
For arbitrary $\bm v\in\mathbb{C}^N,\|\bm v\|_2 = 1$, we extend $\bm v$ to an infinite sequence $v[n], n\in \mathbb Z$ by zero-padding. Then
\begin{align*}
\langle \bm H_N \bm v, \bm v \rangle &= \sum_{m=0}^{N-1}\bm v^*[m]\sum_{n=0}^{N-1} h[m-n]\bm v[n]\\
& = \sum_{m=-\infty}^{\infty} v^*[m]\sum_{n=-\infty}^{\infty} h[m-n] v[n]\\
& = \int_{0}^{1}\left|\widetilde{\bm v}(f)\right|^2\widetilde h(f)df
\end{align*}
where $\widetilde {\bm v}(f) = \sum_{n = 0}^{N-1}\bm v[n]e^{j2\pi f n}$. If $\widetilde h(f)$ is not a constant function of $[0,1]$, we conclude
\begin{align*}
\text{ess}\inf \widetilde h =& \int_{0}^{1}\left|\widetilde{{\bm v}}(f)\right|^2df\cdot\text{ess}\inf \widetilde h < \langle \bm H_N \bm v, \bm v \rangle \\< & \int_{0}^{1}\left|\widetilde{\bm v}(f)\right|^2df\cdot\text{ess}\sup \widetilde h = \text{ess}\sup \widetilde h.
\end{align*}
\cqfd

Theorem \ref{thm:individual eig toeplitz for continuous function} holds trivially when $\widetilde h(f)$ is a constant function since for this case $\bm H_N$ and $\bm C_N$ have the same eigenvalues for all $\bm C_N\in\left\{\widetilde {\bm C}_N, \widehat {\bm C}_N, \overline {\bm C}_N \right\}$ and $N\in\mathbb N$. In what follows, we suppose $\widetilde h(f)$ is not a constant function.
The assumption of absolute summability of the sequence $h[k]$ indicates that its DTFT $\widetilde h(f)$ is continuous on $[0,1]$, and moreover, its partial Fourier sum $S_N(f)$ converges uniformly to $\widetilde h(f)$ on  $[0,1]$ as $N\rightarrow \infty$~\cite{korner1989fourier}. Thus, given $\epsilon>0$, there exists $N_0\in \N$ such that
\begin{align*}
\left|\widetilde h(f) - S_{N-1}(f)\right|\leq \epsilon
\end{align*}
for all $f\in[0,1]$ and $N\geq N_0$. The Ces\`{a}ro sum $\sigma_N(f)$ also converges to $\widetilde h(f)$ uniformly on $[0,1]$ as $N\rightarrow \infty$.

Since the eigenvalues of $\widetilde{\bm C}_N$ and $\widehat{\bm C}_N$ are, respectively, the samples of $S_{N-1}(f)$ and $S_{\left\lfloor\frac{N-1}{2}\right\rfloor}(f)$, we conclude that $\{\widetilde{\bm C}_N\}$ and $\{\widehat{\bm C}_N\}$ are uniformly absolutely bounded. Lemma~\ref{lem:bounds Pearl's eigenvalues} implies that $\{\overline{\bm C}_N\}$ and $\{\bm H_N\}$ are also uniformly absolutely bounded.

We next show $\lim_{N\rightarrow\infty} \max_{l}\lambda_l\left( \bm C_N\right) = \max_{f\in[0,1]}\widetilde h(f)$ for all $\bm C_N\in\left\{\widetilde {\bm C}_N, \widehat {\bm C}_N, \overline {\bm C}_N \right\}$. The extreme value theorem states that $\widetilde h(f)$ must attain a maximum and a minimum each at least once since $\widetilde h(f)\in\mathbb R$ is continuous on $[0,1]$.
Let $$\widehat f: = \argmax_{f}\widetilde h(f)$$
denote any point at which $\widetilde h$ achieves its maximum value.  Also let
$$\widehat l_N : = \argmin_{l\in [N]}\left|\widehat f - \frac{l}{N} \right|$$
denote any closest on-grid point to $\widehat f$.  For arbitrary $\epsilon>0$, by uniform convergence, there exists $N_0$ such that
$$\left|\lambda_{\widehat l_N}\left(\bm C_N \right)- \widetilde h\left(\frac{\widehat l_N}{N}\right)  \right|\leq \epsilon$$
for all $N\geq N_0$.
Noting that $\left|\frac{\widehat l_N}{N} - \widehat f\right|\leq \frac{1}{2N}$ and $\widetilde h$ is continuous on $[0,1]$, there exists $N_1\in \N$ so that
$$\left|\widetilde h\left( \widehat f \right) - \widetilde h\left(\frac{\widehat l_N}{N}  \right)\right|\leq \epsilon$$
when $N\geq N_1$. Thus we conclude
$$\left|\lambda_{\widehat l_N}\left(\bm C_N \right)- \widetilde h\left( \widehat f \right)  \right|\leq 2\epsilon$$
for all $N\geq \max\left\{N_0,N_1\right\}$.
Since $\epsilon$ is arbitrary, $$\lim_{N\rightarrow\infty} \max_{l}\lambda_l\left( \bm C_N\right) = \max_{f\in[0,1]}\widetilde h(f)$$ for all $\bm C_N\in\left\{\widetilde {\bm C}_N, \widehat {\bm C}_N, \overline {\bm C}_N \right\}$. Noting that $\lambda_l\left( \overline{\bm C}_N\right)\leq\lambda_0(\bm H_N)\leq \max_{f\in[0,1]}\widetilde h(f)$, we obtain
$$\lim_{N\rightarrow\infty} \max_{l}\lambda_l\left( \bm H_N\right) = \max_{f\in[0,1]}\widetilde h(f).$$
The asymptotic argument for the smallest eigenvalues can be obtained with a similar approach. It follows from Lemma~\ref{lem:circulantApproxCollective} that $\bm H_N\sim \widehat {\bm C}_N\sim\widetilde {\bm C}_N\sim\overline {\bm C}_N$ and from Szeg\H{o}'s theorem~\eqref{eq:Szego thm} that $$\lim_{N\rightarrow \infty}\frac{1}{N}\sum_{l=0}^{N-1}\vartheta(\lambda_l(\bm{H}_N)) = \int_0^{1}\vartheta(\widetilde{h}(f))df.$$ Finally, the proof of Theorem \ref{thm:individual eig toeplitz for continuous function} is completed by applying Theorem~\ref{thm:uniform to individual for contunuous fuc} with $g = \widetilde{h}$. \cqfd

\subsection{Proof of Theorem \ref{thm:convergence rate band Toeplitz}}

We first note that \cite[Theorem 1.6]{Bogoya2015maximum} provides a similar convergence rate for individual asymptotic  eigenvalue estimates for band Toeplitz matrices by sampling the symbol $\widetilde{h}(f)$ uniformly  in the frequency on $[0,1]$. As illustrated in Appendix~\ref{prf:convergence rate band Toeplitz}, when $h[k]=0$ for all $|k|>r$, the eigenvalues of both $\widetilde {\bm C}_N$ and $\widehat{\bm C}_N$ are equivalent to the DFT samples of $S_{N-1}(f) = \widetilde h(f) =\sum_{k = -r}^{r} h[k]e^{j2\pi fk}$. Thus, \eqref{eq:convergence rate band Toeplitz} for $\widetilde {\bm C}_N$ and $\widehat{\bm C}_N$ follows directly from~\cite[Theorem 1.6]{Bogoya2015maximum}, whose proof involves the quantile function and a refinement of Szeg\H{o}'s asymptotic formula in terms of the number of eigenvalues inside a given interval in~\cite{zizler2002finer}. To keep the paper self-contained and to exhibit an alternative approach, we also prove  \eqref{eq:convergence rate band Toeplitz} for $\widetilde {\bm C}_N$ and $\widehat{\bm C}_N$ by directly bounding the error between the eigenvalues of the band Toeplitz and circulant matrices. The complete proof is given in Appendix \ref{prf:convergence rate band Toeplitz}. We outline the main idea here. Let $[\bm H_N]_{N-r}$  be the $\left(N-r\right)\times \left(N-r\right)$ matrix obtained by deleting the last $r$ columns and the last $r$ rows of $\bm H_N$. Similar notation holds for $[\widetilde{\bm C}_N]_{N-r}$. Note that $[\bm H]_{N-r}$ and $[\widetilde{\bm C}_N]_{N-r}$ have the same eigenvalues when $N>2r$ since $[\bm H]_{N-r}$ is exactly the same as $[\widetilde{\bm C}_N]_{N-r}$. Also $\widehat {\bm C}_N$ is equivalent to $\widetilde {\bm C}_N$ when $N>2r$. We first apply the Sturmian separation theorem for the Toeplitz and circulant matrices to obtain a bound on the distance between $\lambda_l({\bm H}_N)$ and $\lambda_{\rho (l)}(\widetilde{\bm C}_N)$. We then utilize the fact that $\widetilde h(f)$ is Lipschitz continuous to guarantee the closeness between $\lambda_l(\widetilde{\bm C}_N)$ and $\lambda_{l+r}(\widetilde{\bm C}_N)$. Finally, we show $\lambda_l(\widetilde{\bm C}_N)$ is close to $\lambda_l(\overline{\bm C}_N)$ since the Ces\`{a}ro sum and partial Fourier sum converge to the same function in this case.
\cqfd

\subsection{Proof of Theorem \ref{thm:individual eig toeplitz}}
\label{sec:proofremark}

We first provide another condition (which, informally speaking, is weaker than that in Theorem~\ref{thm:uniform to individual for contunuous fuc}) under which the equal distribution of two sequences implies individual asymptotic equivalence.
\begin{thm}
Assume that $b\geq u_{N,0}\geq u_{N,1}\geq \cdots\geq u_{N,N-1}\geq a$ and $b\geq v_{N,0}\geq v_{N,1}\geq \cdots\geq v_{N,N-1}\geq a$. Furthermore, suppose there is a  Riemann integrable function $g(x):[c,d]\rightarrow \bm [a,b]$ such that
\begin{align*}
u_{N,l}, v_{N,l} \in \textup{int}\left(\ess\mathcal{R}(g)\right),\enskip \forall\enskip l\in[N],N\in \N,
\end{align*}
and
\begin{align*}
\lim_{N\rightarrow \infty}\frac{1}{N}\sum_{l=0}^{N-1}\vartheta(u_{N,l}) = \frac{1}{d-c}\int_{c}^d\vartheta(g(x))dx <\infty
\end{align*}
for all $\vartheta$ that are continuous on $[a,b]$. Then the following are equivalent:
\begin{align}
\lim_{N\rightarrow \infty}\frac{1}{N}\sum_{l=0}^{N-1}\left(\vartheta(u_{N,l}) - \vartheta(v_{N,l})\right) = 0;
\label{eq:uniform convergence}\end{align}
\begin{align}
\lim_{N\rightarrow \infty} \max_{l} \left| u_{N,l} - v_{N,l} \right| =0.
\label{eq:individual convergence}\end{align}
\label{thm:uniform to individual}\end{thm}
\noindent{\textbf{Proof}} (of Theorem \ref{thm:uniform to individual}). See Appendix \ref{prf:uniform to individual}. \cqfd

\noindent{\bf Remark.} Theorem~\ref{thm:uniform to individual for contunuous fuc} requires that $g$ is continuous and that the extreme values of the sequences asymptotically converge to the extreme values of $g$ (but meanwhile the extreme values of the sequences can be outside of the range of $g$). Theorem~\ref{thm:uniform to individual} requires the sequences to be strictly inside the range of $g$ (but $g$ can have discontinuities).

Now we are ready to prove Theorem~\ref{thm:individual eig toeplitz}.
If $\widetilde h(f) \equiv C$ is a constant function, then $\lambda_{l}\left(\bm H_N \right) = \lambda_{l}\left(\overline{\bm C}_N \right) = C$ for all $l\in [N]$. Thus Theorem \ref{thm:individual eig toeplitz} holds trivially. On the other hand, suppose that $\widetilde{h}\in L^{\infty}([0,1])$ is not a constant function and
the essential range of $\widetilde h$ is $\left[\essinf \widetilde h,\esssup\widetilde h \right]$. It follows from Lemma~\ref{lem:bounds Pearl's eigenvalues} that $\lambda_l\left( \bm H_N \right), \lambda_l\left( \overline{\bm C}_N \right) \in \mbox{int}\left(\mathcal{R}\left( \widetilde h\right)  \right)$ for all $l\in[N]$ and $N\in\mathbb N$. Using Lemma \ref{lem:circulantApproxCollective} and Szeg\H{o}'s theorem (see \eqref{eq:Szego thm}), the fact that $h[k]$ is square summable together with the fact that $\{\bm H_N\}$ and $\{\overline{\bm C}_N\}$ are uniformly absolutely bounded imply
\[
\lim_{N\rightarrow \infty}\frac{1}{N}\sum_{l=0}^{N-1}\vartheta(\lambda_l(\bm{H}_N)) = \int_0^{1}\vartheta(\widetilde{h}(f))df,
\]
and
\[
\lim_{N\rightarrow \infty}\frac{1}{N}\sum_{l=0}^{N-1}\left( \vartheta(\lambda_l(\bm{H}_N)) -\vartheta(\lambda_l(\overline{\bm C}_{N})) \right) = 0
\]
for all $\vartheta$ that are continuous on $\left[\essinf \widetilde h,\esssup\widetilde h \right]$. Finally, \eqref{eq: individual equiv} follows from Theorem~\ref{thm:uniform to individual} with $g = \widetilde h$, $u_{N,l} = \lambda_l(\bm H_N)$ and $v_{N,l} = \lambda_{\rho(l)}(\overline{\bm C}_N)$.
This completes the proof of Theorem \ref{thm:individual eig toeplitz}. \cqfd

\noindent{\bf Remark.} We note that, as guaranteed by Lemma~\ref{lem:bounds Pearl's eigenvalues} that $\lambda_l(\overline{\bm C}_N)$ and $\lambda_l(\bm H_N)$ are always inside the range of $\widetilde h$ for all $l\in[N]$ and $N\in\N$, Theorem~\ref{thm:uniform to individual} can also be utilized to prove Theorem~\ref{thm:individual eig toeplitz for continuous function} (i.e., \eqref{eq:individual equiv for continuous function}) for $\bm C_N = \overline{\bm C}_N$. However, we cannot apply Theorem~\ref{thm:uniform to individual} for the other two classes of circulant approximations since their eigenvalues can be outside of the range of $\widetilde h$.

\subsection{Proof of Theorem \ref{thm:bounds extreme eigenvalues}}

Our proof of Theorem \ref{thm:bounds extreme eigenvalues} appears in Appendix \ref{prf:bounds extreme eigenvalues}. \cqfd

\section{Simulations}\label{sec:simulations}

In this section, we provide several examples to illustrate our theory. In the legends of Figures~\ref{figure:triangualr Func}--\ref{figure:window Func}, we refer to the circulant approximations $\widetilde {\bm C}_N, \widehat {\bm C}_N$, and $\overline {\bm C}_N$ as Circulant1, Circulant2, and Circulant3, respectively.

\subsection{$h[k] = W\left(\frac{\sin\left(\pi W k  \right)}{\pi k} \right)^2, W = \frac{1}{4}$}

In our first example, the sequence $h[k]$ is absolutely summable and the corresponding symbol
$$
\widetilde h(f) = \text{tri}(\frac{f}{W}) = \left\{\begin{array}{ll}1 -\frac{f}{W}, & 0 \leq f\leq W\\ 1 - \frac{1-f}{W}, & 1-W\leq f\leq 1 \\0, & \mbox{otherwise}\end{array}\right.
$$
is a triangular signal, which is continuous on $[0,1]$. Figure~\ref{figure:triangualr Func}(a) shows $\widetilde h$, Figure~\ref{figure:triangualr Func}(b) shows $\lambda_l(\bm H_N)$,  $\lambda_{\rho(l)}(\widetilde{\bm C}_N)$, $\lambda_{\rho(l)}(\widehat{\bm C}_N)$ and $\lambda_{\rho(l)}(\overline{\bm C}_N)$ for $N = 500$, and  Figure~\ref{figure:triangualr Func}(c) shows $\max_{l\in[N]}\left|\lambda_l(\bm{H}_N) - \lambda_{\rho(l)}({\bm C}_{N})\right|$ against the dimension $N$ for all $\bm C_N\in\left\{\widetilde {\bm C}_N, \widehat {\bm C}_N, \overline {\bm C}_N \right\}$. As guaranteed by Theorem~\ref{thm:individual eig toeplitz for continuous function}, it can be observed in Figure~\ref{figure:triangualr Func}(c) that the individual asymptotic convergence of eigenvalues holds for all $\widetilde {\bm C}_N, \widehat {\bm C}_N$, and $\overline {\bm C}_N$.

\begin{figure*}[t]
\begin{minipage}{0.32\linewidth}
\centering
\includegraphics[width=2.4in]{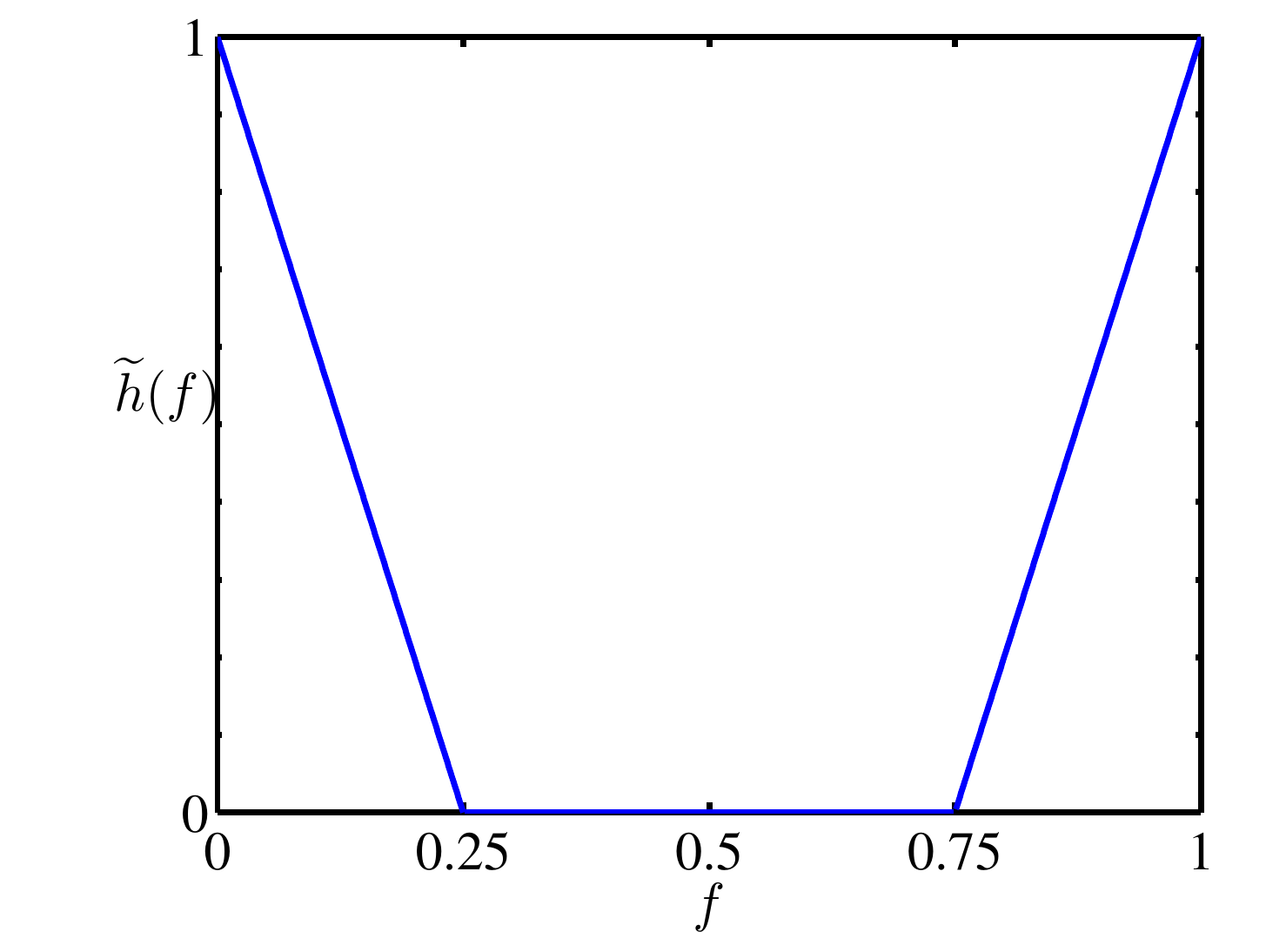}
\centerline{(a)}
\end{minipage}
\hfill
\begin{minipage}{0.32\linewidth}
\centering
\includegraphics[width=2.4in]{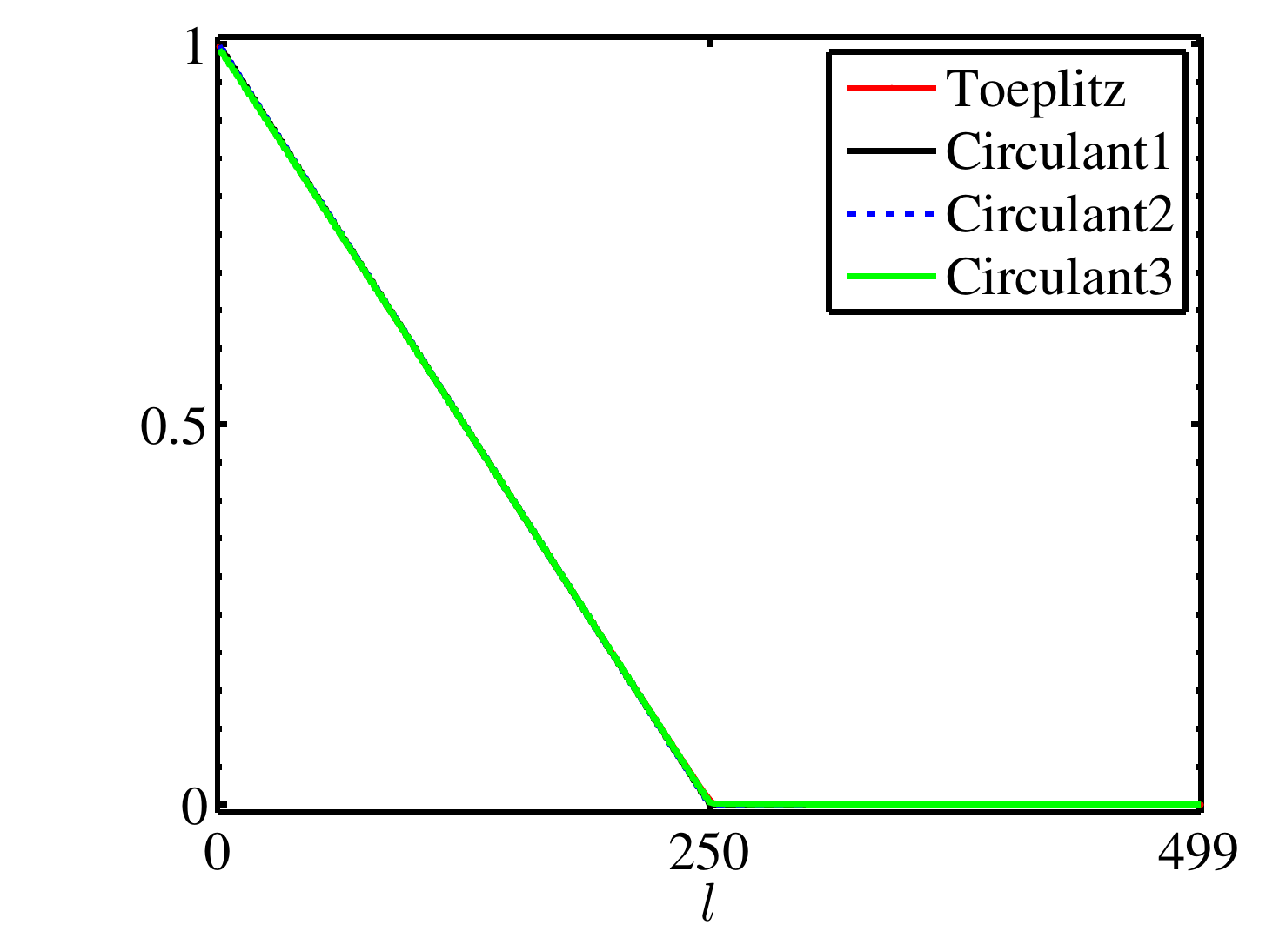}
\centerline{(b)}
\end{minipage}
\hfill
\begin{minipage}{0.32\linewidth}
\centering
\includegraphics[width=2.4in]{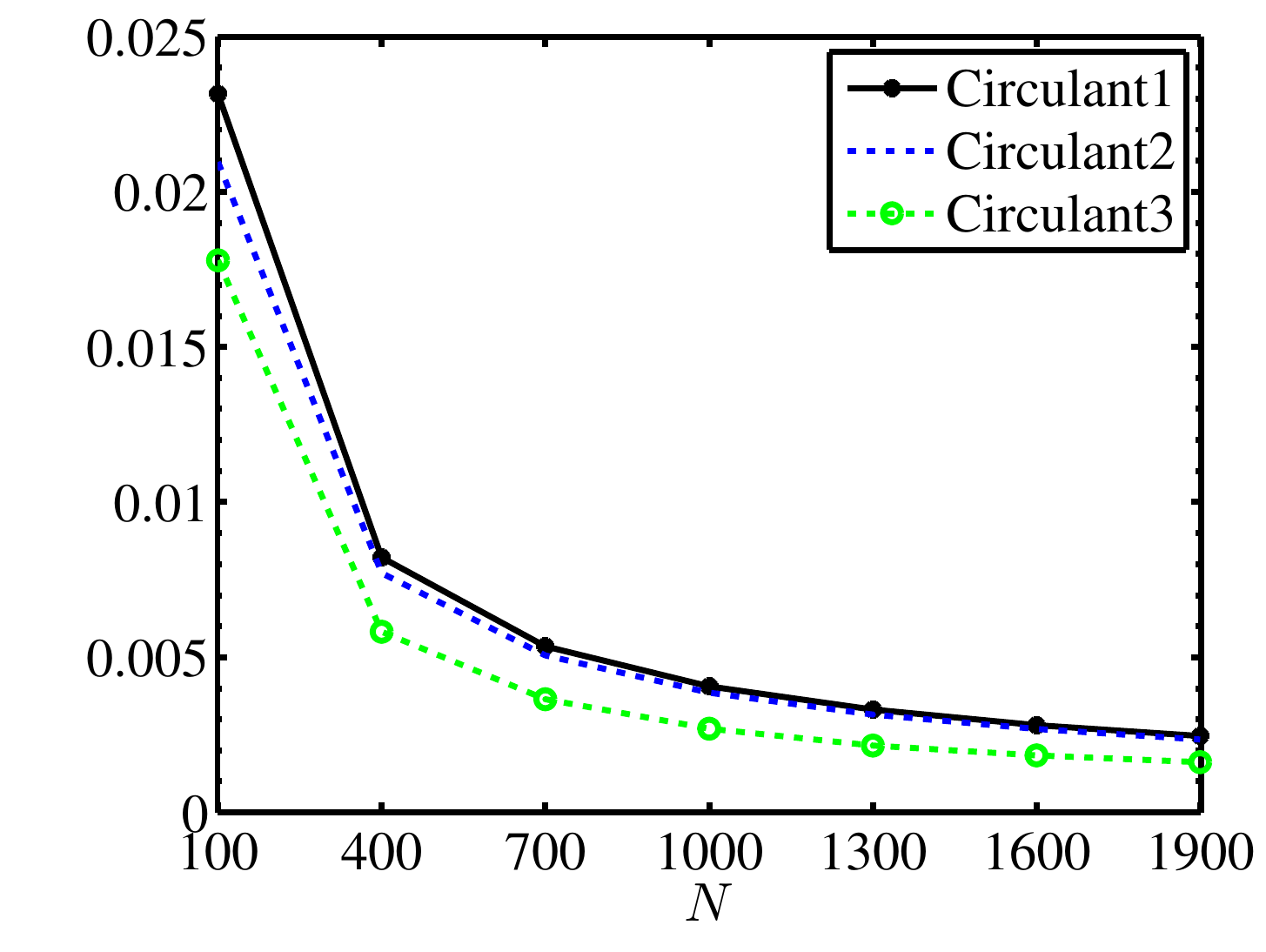}
\centerline{(c)}
\end{minipage}
\caption{ (a) Illustration of a continuous symbol $\widetilde h(f)$. (b) The eigenvalues of the Toeplitz matrix $\bm H_N$ and the circulant approximations $\widetilde {\bm C}_N, \widehat {\bm C}_N$, and $\overline {\bm C}_N$, arranged in decreasing order. Here $N = 500$. (c) A plot of $\max_{l\in[N]}\left|\lambda_l(\bm{H}_N) - \lambda_{\rho(l)}({\bm C}_{N})\right|$ versus the dimension $N$ for all $\bm C_N\in\left\{\widetilde {\bm C}_N, \widehat {\bm C}_N, \overline {\bm C}_N \right\}$.
}\label{figure:triangualr Func}
\end{figure*}

\subsection{$h[k] = \frac{1 + (-1)^k}{j2\pi k}$}\label{sec:example h[k] not summalbe but range is connected}

In this case, the sequence $h[k]$ is not absolutely summable and the symbol
$$
\widetilde h(f)= \left\{\begin{array}{ll}2f, & 0<f\leq \frac{1}{2}\\2f - 1, & \frac{1}{2}<f\leq 1\end{array}\right.$$
is not continuous, but its range is connected. Figure~\ref{figure:discontinuous Func}(a) shows $\widetilde h$, Figure~\ref{figure:discontinuous Func}(b) shows $\lambda_l(\bm H_N)$, $\lambda_{\rho(l)}(\widetilde{\bm C}_N)$, $\lambda_{\rho(l)}(\widehat{\bm C}_N)$ and $\lambda_{\rho(l)}(\overline{\bm C}_N)$ for $N = 500$, and Figure~\ref{figure:discontinuous Func}(c) shows $\max_{l\in[N]}\left|\lambda_l(\bm{H}_N) - \lambda_{\rho(l)}({\bm C}_{N})\right|$ against the dimension $N$ for all $\bm C_N\in\left\{\widetilde {\bm C}_N, \widehat {\bm C}_N, \overline {\bm C}_N \right\}$. It is observed from Figure~\ref{figure:discontinuous Func}(c) that the individual asymptotic convergence of the eigenvalues holds for $\overline {\bm C}_N$---as guaranteed by Theorem~\ref{thm:individual eig toeplitz}---but not for $\widetilde {\bm C}_N$ and $\widehat {\bm C}_N$. Figure~\ref{figure:discontinuous Func}(c) also shows that the errors $\max_{l\in[N]}\left|\lambda_l(\bm{H}_N) - \lambda_{\rho(l)}(\widetilde{\bm C}_{N})\right|$ and $\max_{l\in[N]}\left|\lambda_l(\bm{H}_N) - \lambda_{\rho(l)}(\widehat{\bm C}_{N})\right|$ converge to the size of the Gibbs jump ($\approx0.089$).

\begin{figure*}[t]
\begin{minipage}{0.32\linewidth}
\centering
\includegraphics[width=2.4in]{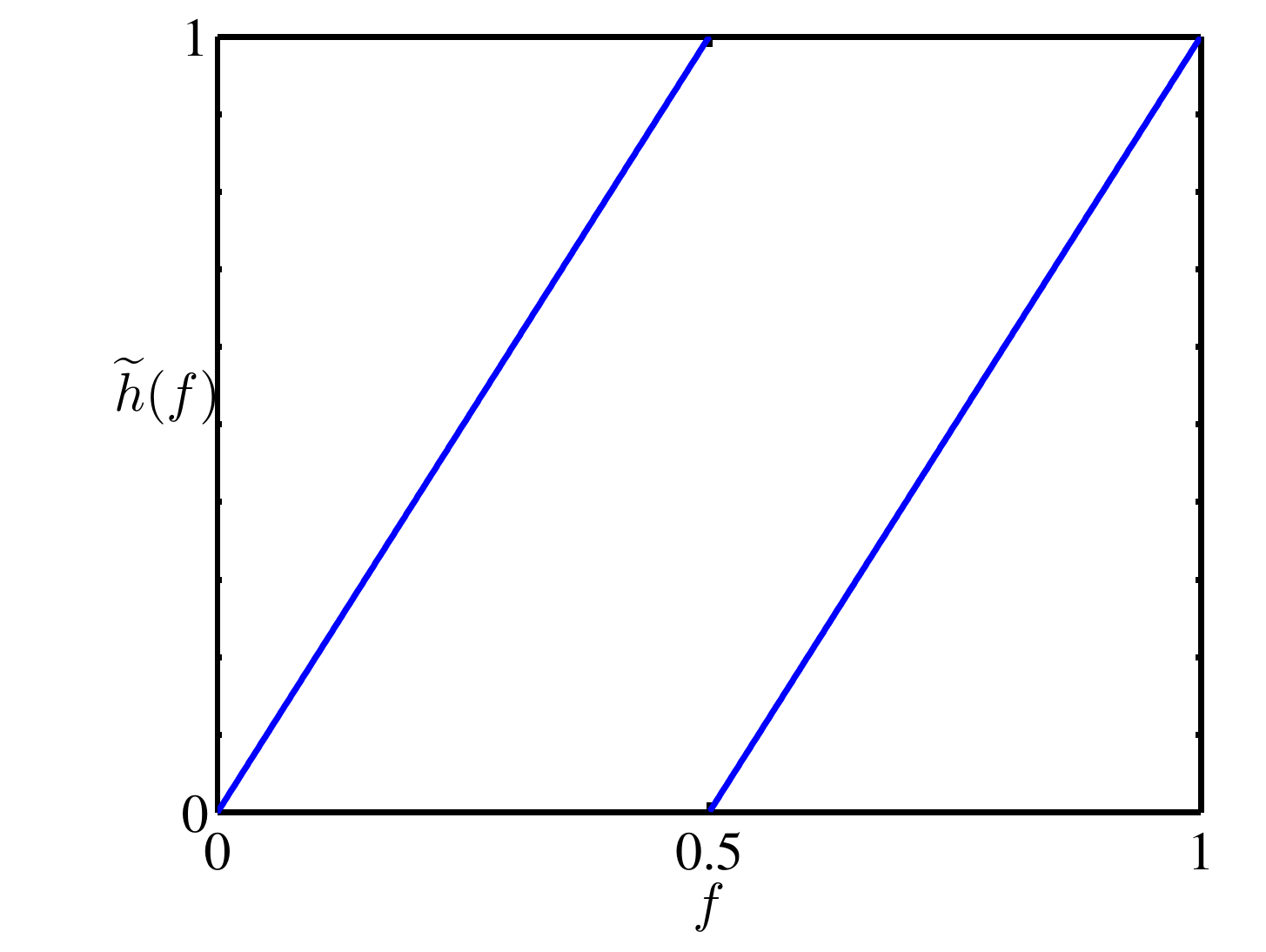}
\centerline{(a)}
\end{minipage}
\hfill
\begin{minipage}{0.32\linewidth}
\centering
\includegraphics[width=2.4in]{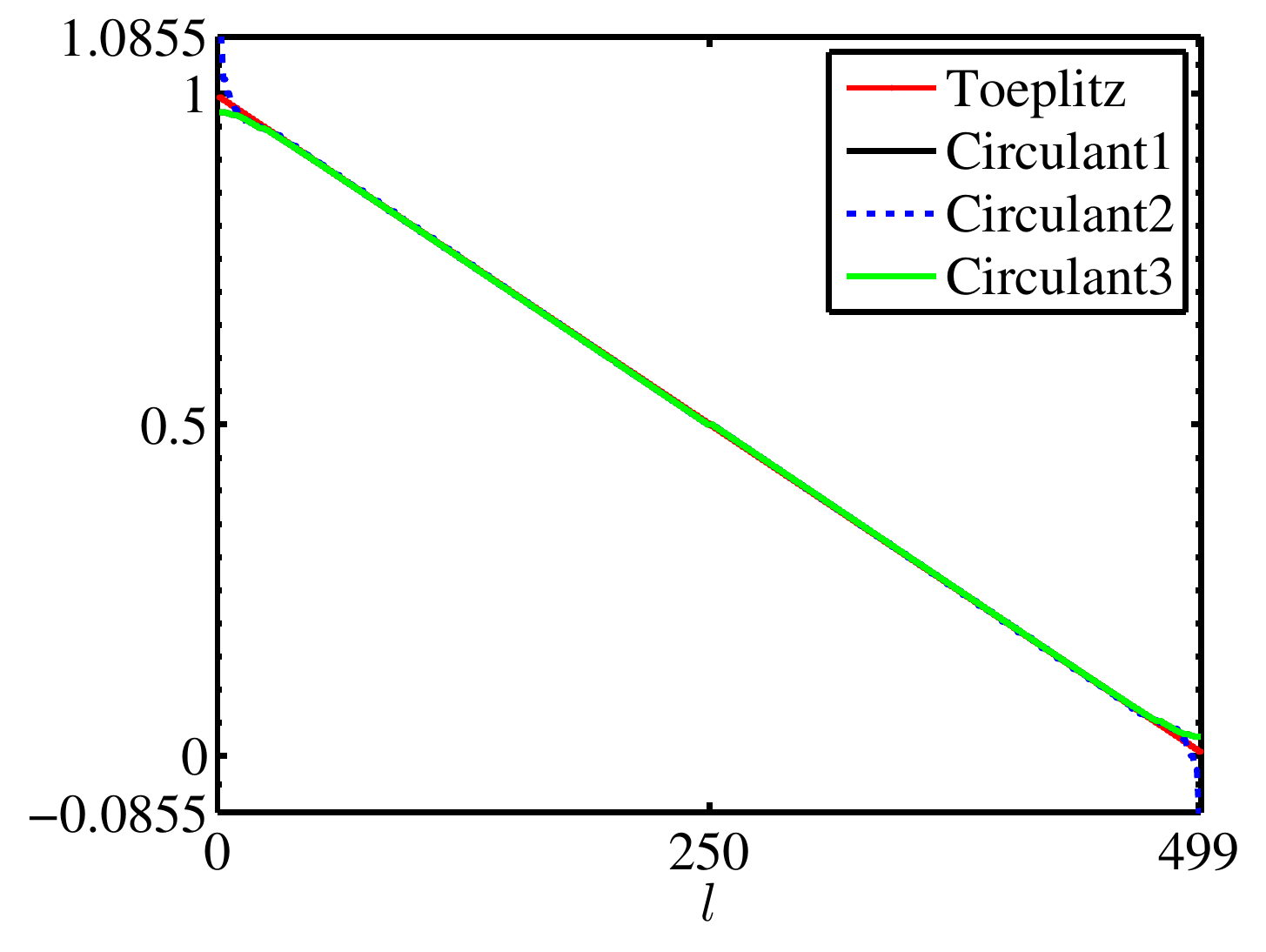}
\centerline{(b)}
\end{minipage}
\hfill
\begin{minipage}{0.32\linewidth}
\centering
\includegraphics[width=2.4in]{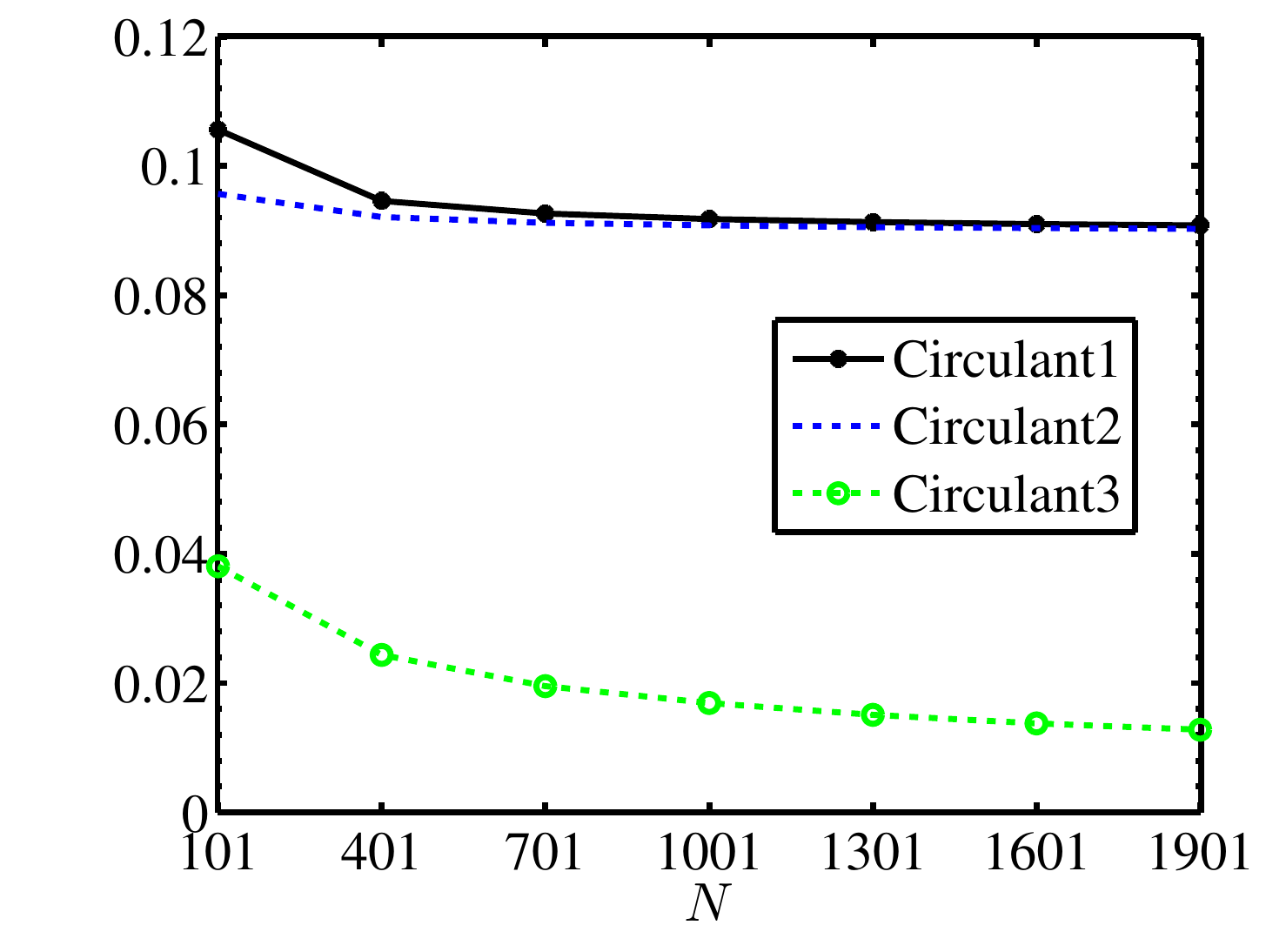}
\centerline{(c)}
\end{minipage}
\caption{(a) Illustration of a discontinuous symbol $\widetilde{h}(f)$. (b) The eigenvalues of the Toeplitz matrix $\bm H_N$ and the circulant approximations $\widetilde {\bm C}_N, \widehat {\bm C}_N$, and $\overline {\bm C}_N$, arranged in decreasing order. Here $N = 500$. (c) A plot of $\max_{l\in[N]}\left|\lambda_l(\bm{H}_N) - \lambda_{\rho(l)}({\bm C}_{N})\right|$ versus the dimension $N$ for all $\bm C_N\in\left\{\widetilde {\bm C}_N, \widehat {\bm C}_N, \overline {\bm C}_N \right\}$.
}\label{figure:discontinuous Func}
\end{figure*}

\subsection{$h[k] = \frac{\sin\left(2\pi W k  \right)}{\pi k}, W = \frac{1}{4}$}

In this example, the sequence $h[k]$ is not absolutely summable and the symbol
$$\widetilde h(f)= \left\{\begin{array}{rc}0, & W<f\leq 1-W,\\1, & \mbox{otherwise},\end{array}\right.$$
is a rectangular window function, which is not continuous and whose range is not connected. Figure~\ref{figure:window Func}(a) shows $\widetilde h$, Figure~\ref{figure:window Func}(b) shows $\lambda_l(\bm H_N)$,  $\lambda_{\rho(l)}(\widetilde{\bm C}_N)$, $\lambda_{\rho(l)}(\widehat{\bm C}_N)$ and $\lambda_{\rho(l)}(\overline{\bm C}_N)$ for $N = 2048$, and  Figure~\ref{figure:window Func}(c) shows $\max_{l\in[N]}\left|\lambda_l(\bm{H}_N) - \lambda_{\rho(l)}({\bm C}_{N})\right|$ against the dimension $N$ for all $\bm C_N\in\left\{\widetilde {\bm C}_N, \widehat {\bm C}_N, \overline {\bm C}_N \right\}$. Figure~\ref{figure:window Func}(c) illustrates that the individual asymptotic convergence of eigenvalues does not hold for the circulant matrices $\widetilde{\bm C}_N$, $\widehat{\bm C}_N$, and $\overline{\bm C}_N$. Indeed, the sequence $h[k]$ does not meet the assumptions in either Theorem~\ref{thm:individual eig toeplitz for continuous function} or Theorem~\ref{thm:individual eig toeplitz}.

\begin{figure*}[t]
\begin{minipage}{0.32\linewidth}
\centering
\includegraphics[width=2.4in]{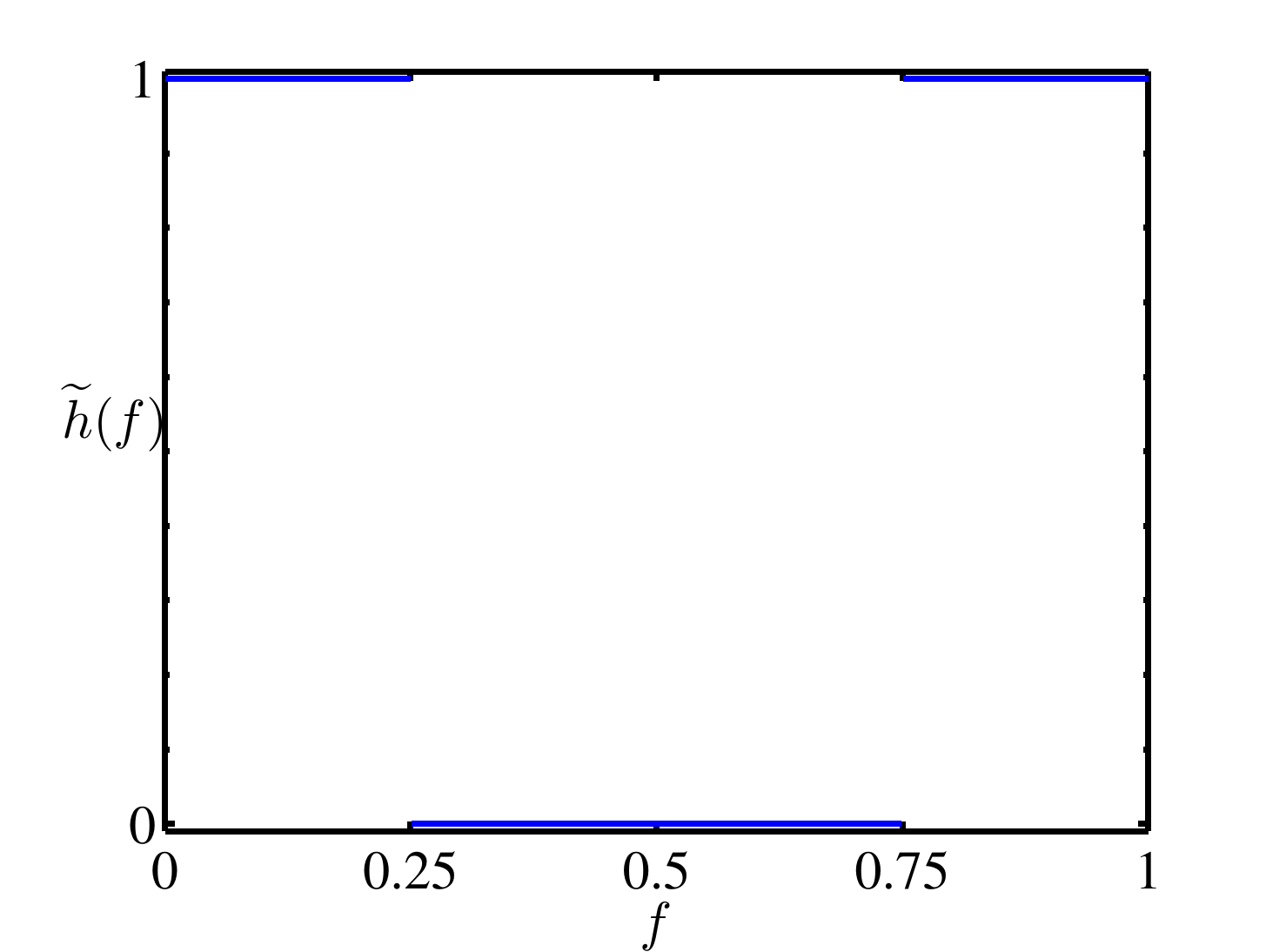}
\centerline{(a)}
\end{minipage}
\hfill
\begin{minipage}{0.32\linewidth}
\centering
\includegraphics[width=2.4in]{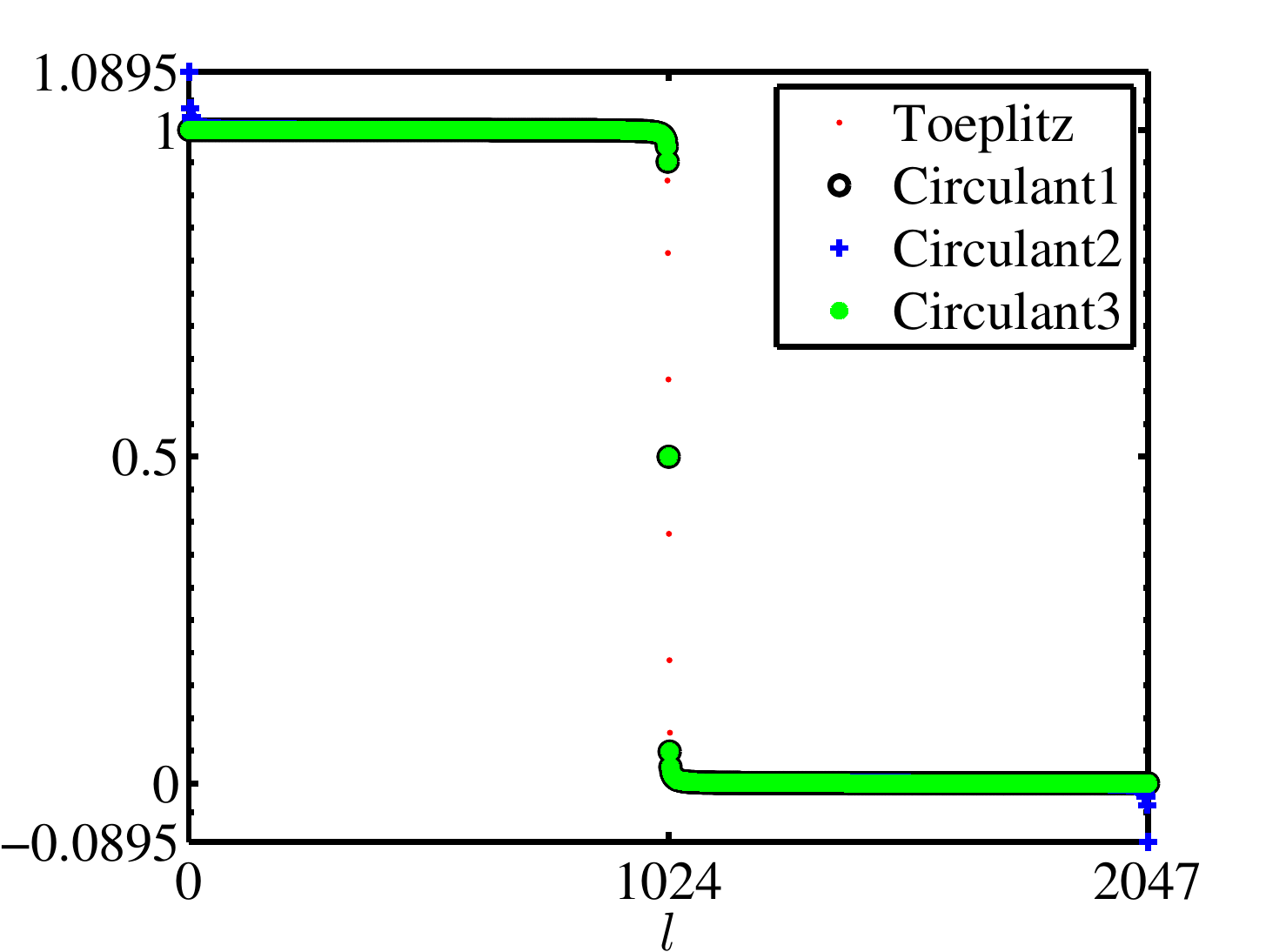}
\centerline{(b)}
\end{minipage}
\hfill
\begin{minipage}{0.32\linewidth}
\centering
\includegraphics[width=2.4in]{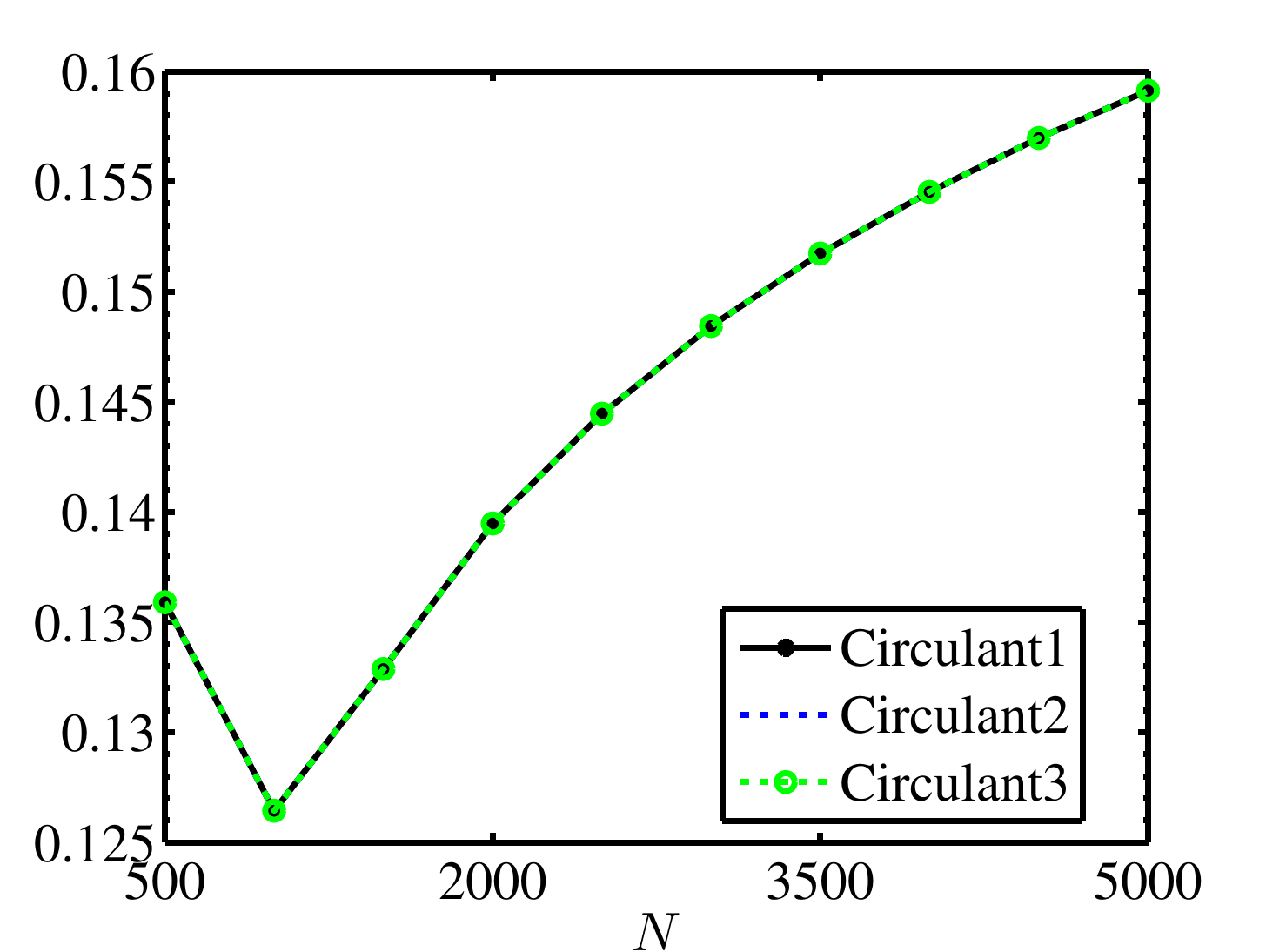}
\centerline{(c)}
\end{minipage}
\vfill
\begin{minipage}{0.32\linewidth}
\centering
\includegraphics[width=2.4in]{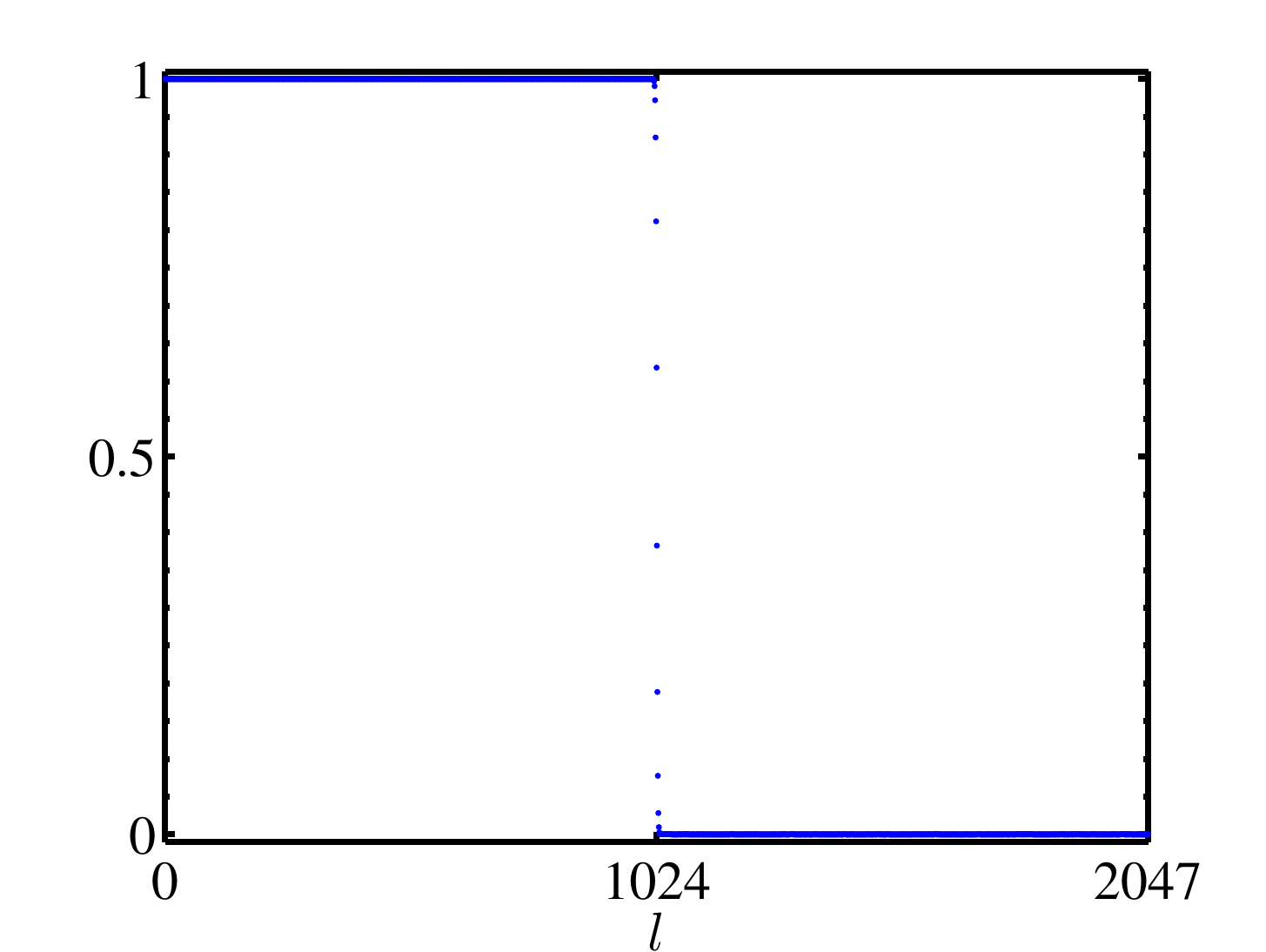}
\centerline{(d)}
\end{minipage}
\hfill
\begin{minipage}{0.32\linewidth}
\centering
\includegraphics[width=2.4in]{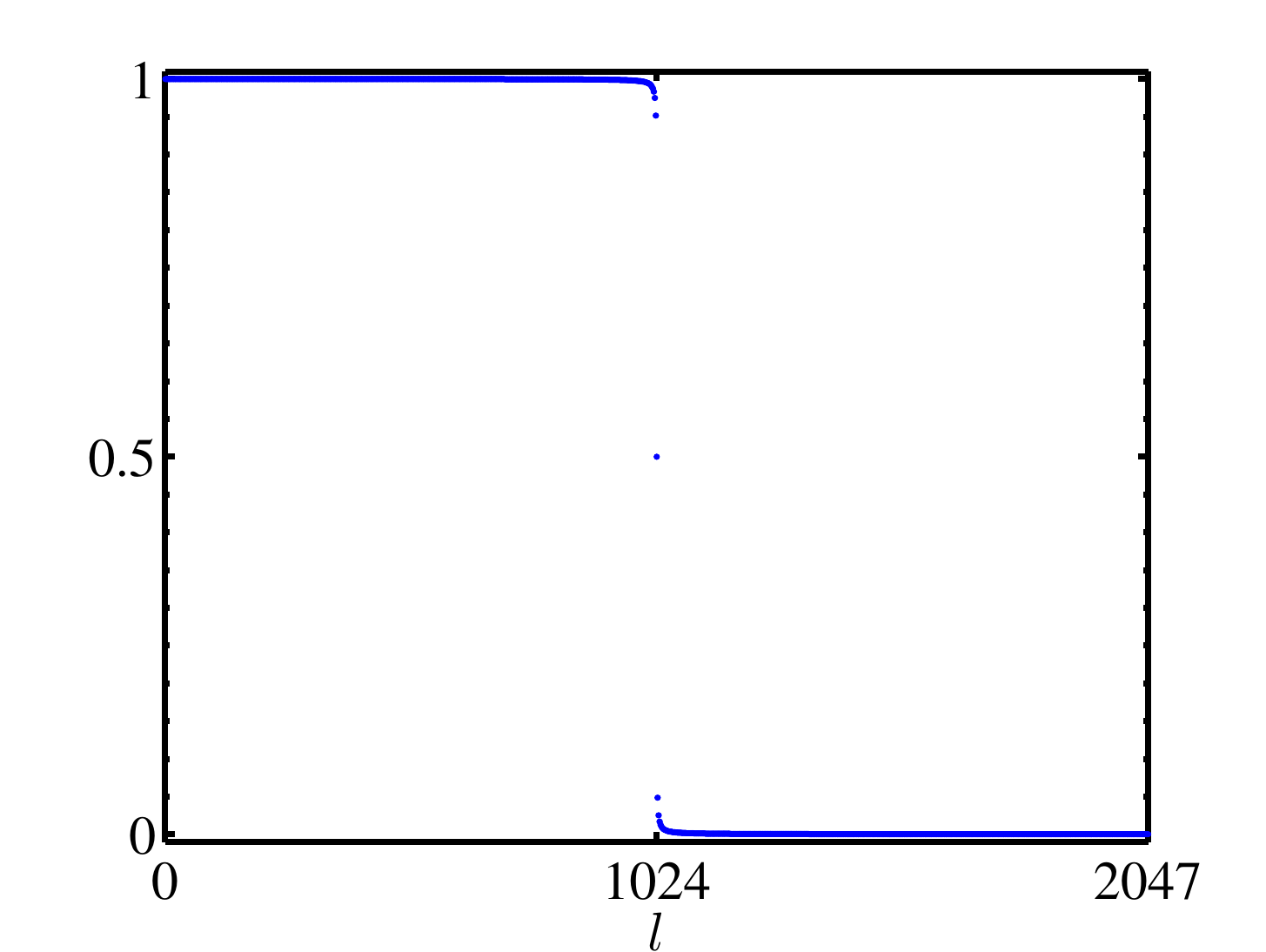}
\centerline{(e)}
\end{minipage}
\hfill
\begin{minipage}{0.32\linewidth}
\centering
\includegraphics[width=2.4in]{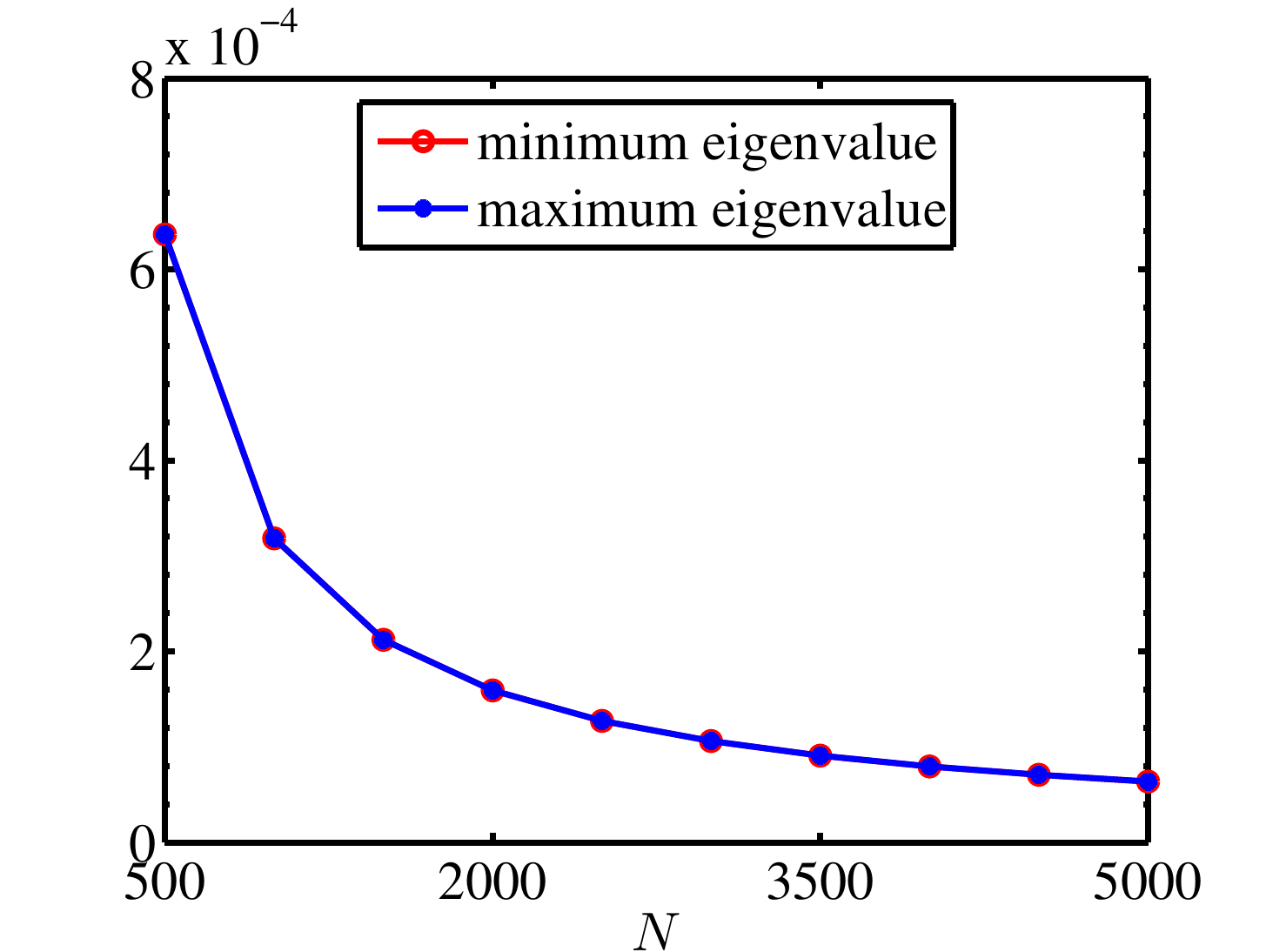}
\centerline{(f)}
\end{minipage}
\caption{(a) Illustration of a discontinuous symbol $\widetilde{h}(f)$ whose range is not connected. (b) The eigenvalues of the Toeplitz matrix $\bm H_N$ and the circulant approximations $\widetilde {\bm C}_N, \widehat {\bm C}_N$, and $\overline {\bm C}_N$, arranged in decreasing order. Here $N = 2048$. (c) A plot of $\max_{l\in[N]}\left|\lambda_l(\bm{H}_N) - \lambda_{\rho(l)}({\bm C}_{N})\right|$ versus the dimension $N$ for all $\bm C_N\in\left\{\widetilde {\bm C}_N, \widehat {\bm C}_N, \overline {\bm C}_N \right\}$. (d) The eigenvalues of the Toeplitz matrix $\bm H_N$. (e) The eigenvalues of the circulant matrix $\overline {\bm C}_N$, arranged in decreasing order. (f) A plot of $\left|\lambda_0(\bm{H}_N) - \lambda_{\rho(0)}(\overline{\bm C}_{N})\right|$ and $\left|\lambda_{N-1}(\bm{H}_N) - \lambda_{\rho(N-1)}(\overline{\bm C}_{N})\right|$ versus the dimension $N$.}\label{figure:window Func}
\end{figure*}

Due to the gap (between $0$ to $1$) in the range of the window function $\widetilde h$, the eigenvalues of $\bm H_N$ and $\overline{\bm C}_N$ have different behavior in the transition region. To better illustrate this, Figures~\ref{figure:window Func}(d) and \ref{figure:window Func}(e), respectively, show $\lambda_l(\bm H_N)$ and $\lambda_{\rho(l)}(\overline{\bm C}_N)$ for $N =2048$. We see that the eigenvalues of the Toeplitz matrix $\bm H_N$ cover the range $[0,1]$ somewhat uniformly, while the eigenvalues of the $\overline{\bm C}_N$ tend to cluster around $0$, $1/2$, and $1$ (there are none near $1/4$ or $3/4$).  The following result formally explains the transition behavior of the eigenvalues of $\bm H_N$.

\begin{lem}\cite{deift2012Spectrum,Slepian78DPSS,zhu2015approximating}
Let $h[k] = \frac{\sin\left(2\pi W k  \right)}{\pi k}$ with $W = \frac{1}{4}$. Fix $\epsilon\in(0,\frac{1}{2})$. Then there exist constants $C_1,C_2$ and $N_1$ such that the distance between any 2 consecutive eigenvalues of $\bm H_N$ inside $(\epsilon, 1-\epsilon)$ is bounded from below by $\frac{C_1}{\ln(N)}$ and from above by $\frac{C_2}{\ln (N)}$; that is
\begin{align*}
\frac{C_1}{\ln(N)} \leq \lambda_l(\bm H_N) - \lambda_{l+1}(\bm H_N) \leq \frac{C_2}{\ln (N)}
\end{align*}
for all $\epsilon\leq\lambda_{l+1}(\bm H_N)\leq \lambda_l(\bm H_N)\leq 1-\epsilon$ and $N\geq N_1$. Also
\begin{align*}
\lambda_{\lfloor \frac{1}{2}N \rfloor-1}\geq \frac{1}{2} \geq \lambda_{\lceil \frac{1}{2}N\rceil}
\end{align*}
for all $N\in\mathbb N$.
\label{lem:transtion on DPSS eigenvalues}
\end{lem}

On the other hand, we have the following result on the eigenvalues of $\overline{\bm C}_N$.

\begin{lem}
Let $h[k] = \frac{\sin\left(2\pi W k  \right)}{\pi k}$ with $W = \frac{1}{4}$. Then
\begin{align*}
\left|\lambda_{l}\left(\overline{\bm C}_N\right) - \frac{1}{2}\right| \left\{\begin{array}{ll}=  0, & l = N/4,3N/4,\\
\geq \alpha,& l\in[N] ~\mbox{and}~ l \neq N/4,3N/4,  \end{array}\right.
\end{align*}
with $\alpha= 0.4$ if $N$ is a multiple of $4$.
\label{lem:Pearl circulant for window function}\end{lem}

\noindent{\textbf{Proof.}} See Appendix \ref{prf:Pearl circulant for window function}. \cqfd

With more sophisticated analysis, we believe that the above result could be improved to $\alpha \approx 0.45$. This is suggested by Figure~\ref{figure:window Func}(e).

Combining Lemmas~\ref{lem:transtion on DPSS eigenvalues} and \ref{lem:Pearl circulant for window function}, we conclude  that $\max_{l\in[N]}\left|\lambda_l(\bm{H}_N) - \lambda_{\rho(l)}(\overline{\bm C}_N)\right|$ approaches $\approx 0.4$ as $N\rightarrow \infty$ and $N$ is a multiple of $4$.

Finally, Figure \ref{figure:window Func}(f) plots $\left|\lambda_0(\bm{H}_N) - \lambda_{\rho(0)}(\overline{\bm C}_{N})\right|$ and $\left|\lambda_{N-1}(\bm{H}_N) - \lambda_{\rho(N-1)}(\overline{\bm C}_{N})\right|$ against the dimension $N$. As can be observed, the largest and smallest eigenvalues of $\overline{\bm C}_N$ converge to the largest and smallest eigenvalues of $\bm H_N$, respectively. This is as guaranteed by Theorem~\ref{thm:bounds extreme eigenvalues}.

\section{Conclusions}\label{sec:conclusions}

It is well known that any sequence of uniformly bounded Hermitian Toeplitz matrices is asymptotically equivalent to certain sequences of circulant matrices derived from the Toeplitz matrices. We have provided conditions under which the asymptotic equivalence of the matrices implies the individual asymptotic convergence of the eigenvalues. Our results suggest that instead of directly computing the eigenvalues of a Toeplitz matrix, one can compute a fast spectrum approximation using the FFT. This is long known, but we provide new guarantees for the asymptotic convergence of the individual eigenvalues. Some numerical examples have demonstrated the dependence of the convergence behavior on the properties of the symbol of the Toeplitz matrix. An interesting question would be whether it is possible to extend our analysis to general (non-Hermitian) Toeplitz matrices, along the lines of the Avram-Parter theorem~\cite{avram1988bilinear,parter1986distribution}. In addition, it would also be of interest to extend our analysis to the asymptotic equivalence of block Toeplitz and block circulant matrices.

\section*{Acknowledgements}

The authors would like to thank the reviewers for their constructive comments and helpful suggestions.

\appendices
\section{Proof of Lemma \ref{lem:circulantApproxCollective}}\label{prf:circulantApproxCollective}
It follows from the definition of $\widehat {\bm C}_N$ that
\begin{align*}
&\left\|\bm H_N - \widehat{\bm C}_N \right\|_F^2 \\ =& \sum_{k=1}^{\left\lfloor\frac{N-1}{2}\right\rfloor}k\left(\left|h[k] - h[-N+k] \right|^2 + \left|h[-k] - h[N-k] \right|^2\right) \\ &+ \sum_{k = \left\lfloor\frac{N-1}{2}\right\rfloor + 1}^{\lfloor N/2 \rfloor} k \left(\left|h[k]\right|^2 + \left|h[-k]\right|^2 \right)\\
 \leq &\sum_{k=1}^{\lfloor N/2 \rfloor} 2k \left(\left|h[k]\right|^2 + \left|h[-k]\right|^2 + \left|h[N-k]\right|^2 + \left|h[k-N]\right|^2 \right)\\
\leq & \sum_{k=1}^{N-1} 2k \left(\left|h[k]\right|^2 + \left|h[-k]\right|^2  \right).
\end{align*}
Fix $\epsilon>0$. By assumption that the sequence $h[k]$ is square summable, there exists $N_0$ such that
\begin{align*}
\sum_{k=N_0}^\infty \left|h[k]\right|^2 + \left|h[-k]\right|^2 \leq \epsilon.
\end{align*}
Thus we have
\begin{align*}
&\frac{1}{N}\left\|\bm H_N - \widehat{\bm C}_N \right\|_F^2\\  \leq & \frac{1}{N}\sum_{k=1}^{N_0-1} 2k \left(\left|h[k]\right|^2 + \left|h[-k]\right|^2  \right) \\ &+ \frac{1}{N}\sum_{k=N_0}^{N} 2k \left(\left|h[k]\right|^2 + \left|h[-k]\right|^2  \right)\\
 \leq &\frac{1}{N}\sum_{k=1}^{N_0-1} 2k \left(\left|h[k]\right|^2 + \left|h[-k]\right|^2  \right) + 2\sum_{k=N_0}^{N} \left(\left|h[k]\right|^2 + \left|h[-k]\right|^2  \right)\\
\leq & \epsilon + 2\epsilon = 3\epsilon
\end{align*}
when $N\geq\max\left\{N_0,N_1\right\}$ with $N_1\geq \sum_{k=1}^{N_0-1} 2k \left(\left|h[k]\right|^2 + \left|h[-k]\right|^2  \right)/\epsilon$. Since $\epsilon$ is arbitrary, we obtain
\begin{align*}
\lim_{N\rightarrow \infty} \frac{1}{N}\left\|\bm H_N - \widehat{\bm C}_N \right\|_F^2 = 0.
\end{align*}
Noting that $\{\bm H_N\}$ and $\{\widehat {\bm C}_N\}$ are uniformly absolutely bounded by assumption, we conclude $\bm H_N\sim \widehat {\bm C}_N$. The proofs of $\bm H_N\sim \widetilde {\bm C}_N$ and $\bm H_N\sim \overline {\bm C}_N$ follow from the same approach. Invoking Theorem~\ref{thm:equivalence of matrices} completes the proof. \cqfd

\section{Proof of Theorem \ref{thm:uniform to individual for contunuous fuc}}\label{prf:uniform to individual for contunuous fuc}
Set \begin{align}
F_g(\alpha) &:= \frac{1}{d-c}\mu\left\{x\in[c,d]: g(x)\leq \alpha \right\},\label{eq:def F_g}\\
F_{u_{N}}(\alpha) &:= \frac{1}{N}\#\left\{l\in[N], u_{N,l}\leq \alpha \right\},\nonumber\\
F_{v_{N}}(\alpha) &:= \frac{1}{N}\#\left\{l\in[N], v_{N,l}\leq \alpha \right\}.\nonumber
\end{align}
Here, $\mu(E)$ is the Lebsegue measure of a subset $E\in \R$.

Definition~\ref{def:Weyl equal distribution} states that the sequences $\{ \{u_{N,l}\}_{l\in [N]} \}_{N=1}^{\infty}$ and  $\{ \{v_{N,l}\}_{l\in [N]} \}_{N=1}^{\infty}$ are asymptotically equally distributed if
\begin{align*}\lim_{N\rightarrow \infty}\frac{1}{N}\sum_{l=0}^{N-1}\left(\vartheta\left(u_{N,l}\right) - \vartheta\left(v_{N,l}\right)\right) = 0
 \end{align*}
for all $\vartheta$ that are continuous on $[a,b]$. Here $[a,b]$ is the smallest interval that covers the sequences $\{\{ u_{N,l}\}_{l\in [N]}\}_{N=1}^\infty$ and $\{\{ v_{N,l}\}_{l\in [N]}\}_{N=1}^\infty$.

Trench~\cite{trench2012elementary} strengthens this definition by showing the following result.
\begin{lem}\cite[Asymptotcially (absolutely) equal distribution]{trench2012elementary}
Assume that $b\geq u_{N,0}\geq u_{N,1}\geq \cdots\geq u_{N,N-1}\geq a$ and $b\geq v_{N,0}\geq v_{N,1}\geq \cdots\geq v_{N,N-1}\geq a$. The following are equivalent:
\begin{enumerate}
\item
$
\lim_{N\rightarrow \infty}\frac{1}{N}\sum_{l=0}^{N-1}\left(\vartheta\left(u_{N,l}\right) - \vartheta\left(v_{N,l}\right)\right) = 0$ for all $\vartheta$ that are continuous on $[a,b]$;
\item $\lim_{N\rightarrow \infty}\frac{1}{N}\sum_{l=0}^{N-1}\left|\vartheta\left(u_{N,l}\right) - \vartheta\left(v_{N,l}\right)\right| = 0$ for all $\vartheta$ that are continuous on $[a,b]$.
\end{enumerate}
\label{lem: absolute uniform converg}\end{lem}
\noindent Here the sequences  $\{ \{u_{N,l}\}_{l\in [N]} \}_{N=1}^{\infty}$ and $\{ \{v_{N,l}\}_{l\in [N]} \}_{N=1}^{\infty}$ are said to be {\it absolutely asymptotically equally distributed}~\cite{trench2012elementary} if
\begin{align*}\lim_{N\rightarrow \infty}\frac{1}{N}\sum_{l=0}^{N-1}\left|\vartheta\left(u_{N,l}\right) - \vartheta\left(v_{N,l}\right)\right| = 0
 \end{align*}
for all $\vartheta$ that are continuous on $[a,b]$.

Viewing $g: [c,d]\rightarrow \R$ as a random variable, in probabilistic language, $F_g$ is the cumulative distribution function (CDF) associated to $g$. Also $F_{u_N}$ and $F_{v_N}$ can be viewed as the CDF of the discrete random variables $\bm u_{N}:\left\{0,1,\ldots,N-1 \right\}\rightarrow \R$ defined by $\bm u_{N}(l) = u_{N,l}$ and $\bm v_{N}:\left\{0,1,\ldots,N-1 \right\}\rightarrow \R$ defined by $\bm v_{N}(l) = v_{N,l}$, respectively. It is well known that the CDF of a random variable is right continuous and non-decreasing. The following result, known as the Portmanteau Lemma, gives two equivalent descriptions of weak convergence in terms of the CDF and the means of the random variables.
\begin{lem}\cite[Portmanteau Lemma]{van2000asymptotic}
The following are equivalent:
\begin{enumerate}
\item$\lim_{N\rightarrow \infty}\frac{1}{N}\sum_{l=0}^{N-1}\vartheta(u_{N,l}) = \frac{1}{d-c}\int_{c}^d\vartheta(g(x))dx$, for all bounded, continuous functions
     $\vartheta$;
\item $\lim_{N\rightarrow \infty}F_{u_{N}}(\alpha) = F_g(\alpha)$
for every point $\alpha$ at which $F_g$ is continuous.
\end{enumerate}
\label{lem:convergence cdf}\end{lem}

Despite the fact that $F_g(\alpha)$ is right continuous and non-decreasing everywhere, some stronger results about $F_g(\alpha)$ can be obtained by utilizing the fact that $g$ is continuous on $[c,d]$.
\begin{lem}
Let $F_g(\alpha)$ be defined as in~\eqref{eq:def F_g}. Then $F_g(\alpha)$ is strictly increasing on $\mathcal R(g)$, i.e., for every $\alpha \in \mbox{int}\left(\mathcal R(g)\right)$, there exists $\epsilon>0$ such that, for each pair $(\alpha_1,\alpha_2)$ satisfying $$\min_{x\in[c,d]}g(x)\leq \alpha -\epsilon<\alpha_1<\alpha<\alpha_2<\alpha+\epsilon\leq \max_{x\in[c,d]}g(x),$$
we have
$$F_g(\alpha_1)<F_g(\alpha)<F_g(\alpha_2).$$
\label{lem:F stricitly increasing for continuous func}\end{lem}
\noindent{\textbf{Proof}} (of Lemma \ref{lem:F stricitly increasing for continuous func}). Since $g(x):[c,d]\rightarrow \R$ is continuous, there exists $\epsilon$ such that $\left(\alpha - \epsilon, \alpha + \epsilon \right) \subset \mathcal{R}(g)$ for $\alpha\in\mbox{int}\left(\mathcal  R(g)\right)$. Let $\alpha_1$ be an arbitrary value such that $\alpha -\epsilon<\alpha_1<\alpha$ and let $\alpha_1' = \frac{\alpha + \alpha_1}{2} \in \mathcal{R}(g)$. Noting that $g$ is continuous, we have
$$\mu\left\{x\in[c,d]: \left|g(x) - \alpha_1'\right|<\frac{\alpha-\alpha_1}{2}  \right\} >0 .$$
Thus, we obtain
\begin{align*}F_g(\alpha) - F_g(\alpha_1) &= \frac{1}{d-c}\mu\left\{ x\in[c,d]: \alpha_1<g(x)\leq \alpha  \right\}\\&\geq \frac{1}{d-c}\mu\left\{ x\in[c,d]: \alpha_1<g(x)< \alpha  \right\}>0.
\end{align*}
Similarly, we have $F_g(\alpha)< F_g(\alpha_2)$ for $\alpha<\alpha_2<\alpha+\epsilon$. \cqfd
%

We are now ready to prove the main part. First we show that \eqref{eq:individual convergence for continuous func} implies \eqref{eq:uniform convergence for continuous func}. Fix $\vartheta$ being some continuous function on $[a,b]$ and $\epsilon>0$. The Weierstrass approximation theorem states that there exists a polynomial $p$ on $[a,b]$ such that
\begin{align*}
\left|\vartheta(t) - p(t)\right| \leq \frac{\epsilon}{3}
\end{align*}
for all $t\in[a,b]$. Since $p$ is a polynomial, there exists a constant $C$ such that
$$\left| p(t_2) - p(t_1) \right|\leq C \left| t_2 - t_1\right|$$
for any $a\leq t_1\leq t_2 \leq b$. Also \eqref{eq:individual convergence for continuous func} implies that there exists an $N_0\in\N$ such that
$$\left|u_{N,l} - v_{N,l} \right|\leq \frac{\epsilon}{3C},\enskip \forall\enskip l\in[N]$$
for all $N\geq N_0$. Therefore, we have
\begin{align*}
&\left|\vartheta(u_{N,l}) - \vartheta(v_{N,l}) \right|\\ \leq & \left|\vartheta(u_{N,l}) - p(u_{N,l}) \right| + \left|p(u_{N,l}) - p(v_{N,l}) \right| \\
& + \left|\vartheta(v_{N,l}) - p(v_{N,l}) \right|\\
 \leq & \frac{\epsilon}{3} + C\frac{\epsilon}{3C} + \frac{\epsilon}{3} = \epsilon
\end{align*}
for all $l\in[N]$ and $N\geq N_0$. Thus
\begin{align*}\left|\frac{1}{N}\sum_{l=0}^{N-1}\left(\vartheta\left(u_{N,l}\right) - \vartheta\left(v_{N,l}\right)\right)\right|\leq &\frac{1}{N}\sum_{l=0}^{N-1}\left|\vartheta(u_{N,l}) - \vartheta(v_{N,l}) \right|\\ \leq & \epsilon
\end{align*}
for all $N\geq N_0$. Since $\epsilon$ is arbitrary, this implies \eqref{eq:uniform convergence for continuous func}.

Now let us show that \eqref{eq:uniform convergence for continuous func} implies \eqref{eq:individual convergence for continuous func}. We prove the statement \eqref{eq:uniform convergence for continuous func} $\Rightarrow$ \eqref{eq:individual convergence for continuous func} by contradiction. Suppose \eqref{eq:individual convergence for continuous func} is not true, i.e., there exists an increasing sequence $\left\{M_{k'}\right\}_{k'=1}^{\infty}$  and $\epsilon_1>0$ such that
$$\max_{l\in[M_{k'}]}\left| u_{M_{k'},l} - v_{M_{k'},l} \right|\geq 2\epsilon_1$$
for all $k'\geq 1$. Let $l_{k'} = \arg\max_{l\in[M_{k'}]}\left| u_{M_{k'},l} - v_{M_{k'},l} \right|$ denote any point at which $\left| u_{M_{k'},l} - v_{M_{k'},l} \right|$ achieves its maximum, which implies  $\left| u_{M_{k'},l_{k'}} - v_{M_{k'},l_{k'}} \right|\geq 2\epsilon_1$. Without loss of generality, we suppose $\{M_{k''},l_{k''}\}_{k''=1}^\infty$ is a subsequence of $\{M_{k'},l_{k'}\}_{k'=1}^\infty$ such that
$u_{M_{k''},l_{k''}}\leq  v_{M_{k''},l_{k''}}$, i.e., $u_{M_{k''},l_{k''}}\leq  v_{M_{k''},l_{k''}} - 2\epsilon_1$.

Note that $u_{M_{k''},l_{k''}}\leq  v_{M_{k''},l_{k''}} - 2\epsilon_1$ together with $\lim_{N\rightarrow \infty}v_{N,0} = \max_{x\in[c,d]}g(x)$ implies $u_{M_{k''},l_{k''}}< \max_{x\in[c,d]}g(x)$ for all sufficiently large $k''$. Thus we let $\{M_{k},l_{k}\}_{k=1}^\infty$ be a subsequence of $\{M_{k''},l_{k''}\}_{k''=1}^\infty$ such that $u_{M_k,l_k}< \max_{x\in[c,d]}g(x)$. By assumption that
\begin{align*}
\lim_{N\rightarrow \infty} u_{N,N-1} = \lim_{N\rightarrow \infty} v_{N,N-1} = \min_{x\in[c,d]}g(x),
\end{align*}
there exist $k_0\in \N$ and $\alpha_k\in \mbox{int}\left(\mathcal{R}(g)\right)$ such that
$$ 0\leq \alpha_k - u_{M_k,l_k}< \frac{\epsilon_1}{2}$$
and $F_g$ is continuous\footnote{If $F_g$ is continuous at $u_{M_k,l_k}$, then we can pick $\alpha_k = u_{M_k,l_k}$. Otherwise,
since $F_g$ is continuous almost everywhere, there always exists an $\alpha_k$ that is close to $u_{M_k,l_k}$ and such that $F_g$ is continuous at $\alpha_k$.} at $\alpha_k$ for all $k\geq k_0$.  Noting that $F_g$ is right continuous everywhere and strictly increasing at all $\alpha_{k}$ (which is shown in Lemma~\ref{lem:F stricitly increasing for continuous func}), there exist $\epsilon_2(\alpha_k)>0$ (which depends on $\alpha_k$) and $\epsilon_3>0$  (which is independent of $\alpha_k$) such that $\epsilon_2(\alpha_k)\leq \frac{\epsilon_1}{2}$, $F_g$ is continuous at $\alpha_{k} + \epsilon_2(\alpha_k)$, and
\begin{align}
F_g(\alpha_{k} + \epsilon_2(\alpha_k)) \geq F_g(\alpha_{k}) + 3\epsilon_3.
\label{eq:F_u strictly continuous for continuous func}\end{align}
 Lemma~\ref{lem:convergence cdf} indicates that
$$\lim_{M_k\rightarrow \infty}F_{u_{M_k}}(\alpha) = F_g(\alpha)$$
for every point $\alpha$ at which $F_g$ is continuous.
Thus there exist $k_1\in \N, k_1\geq k_0$ such that
\begin{equation}\begin{split}
\left| F_{u_{M_k}}\left(\alpha_k\right) - F_g\left(\alpha_k \right)\right| &< \epsilon_3,\\ \left| F_{u_{M_k}}\left(\alpha_k + \epsilon_2(\alpha_k)\right) - F_g\left(\alpha_k + \epsilon_2(\alpha_k) \right)\right| &< \epsilon_3
\end{split}
\label{eq:cdf equivalent for continuous func}\end{equation}
for all $k\geq k_1$.
Thus, we have
\begin{align*}
&F_{u_{M_k}}(\alpha_k + \epsilon_2(\alpha_k)) -F_{u_{M_k}}(\alpha_k)
\\ = &  F_{u_{M_k}}(\alpha_k + \epsilon_2(\alpha_k))  - F_{g}(\alpha_k + \epsilon_2(\alpha_k))  +  F_{g}(\alpha_k + \epsilon_2(\alpha_k) ) \\ & - F_{g}(\alpha_k ) +  F_{g}(\alpha_k)  - F_{u_{M_k}}(\alpha_k)  \\ \geq &  F_{g}(\alpha_k + \epsilon_2(\alpha_k) )  - F_{g}(\alpha_k) - \left| F_{g}(\alpha_k)  - F_{u_{M_k}}(\alpha_k)\right|\\& -\left| F_{u_{M_k}}(\alpha_k + \epsilon_2(\alpha_k))  - F_{g}(\alpha_k + \epsilon_2(\alpha_k)) \right|\\
 \geq & 3\epsilon_3 -\epsilon_3 -\epsilon_3 = \epsilon_3
\end{align*}
for all $k\geq k_1$, where the last line follows from \eqref{eq:F_u strictly continuous for continuous func} and \eqref{eq:cdf equivalent for continuous func}. Noting that the above equation is equivalent to
$$\frac{1}{M_k}\#\left\{l\in[M_k], \alpha_k<u_{M_k,l}\leq \alpha_k + \epsilon_2(\alpha_k)  \right\}\geq \epsilon_3,$$
we have
\begin{align*}&\frac{1}{M_k}\#\left\{l\in[M_k], u_{M_k,l_k}<u_{M_k,l}\leq \alpha_k + \epsilon_2(\alpha_k)  \right\} \\\geq &  \frac{1}{M_k}\#\left\{l\in[M_k], \alpha_k<u_{M_k,l}\leq \alpha_k + \epsilon_2(\alpha_k)  \right\}\geq \epsilon_3.
\end{align*}
Thus, we obtain
$$0\leq u_{M_k,l_k - \lceil  \epsilon_3 M_k \rceil} - u_{M_k, l_k} \leq \alpha_k + \epsilon_2(\alpha_k) - u_{M_k,l_k}  \leq  \epsilon_1,$$
which implies
\begin{align*}&v_{M_k, l_k } - u_{M_k,l_k - \lceil  \epsilon_3 M_k \rceil}\\ = & v_{M_k, l_k } - u_{M_k, l_k } + u_{M_k, l_k } - u_{M_k,l_k - \lceil  \epsilon_3 M_k \rceil}\\
 \geq & 2\epsilon_1 - \epsilon_1 \geq \epsilon_1.
\end{align*}
Now taking $\vartheta(t) = t$, we obtain
\begin{align*}&\frac{1}{M_k}\sum_{l=0}^{M_k-1}\left|\vartheta(u_{M_k,l}) - \vartheta(v_{M_k,l})\right| \\ \geq &\frac{1}{M_k}\sum^{l_k}_{l=l_k - \lceil  \epsilon_3 M_k \rceil}\left|\vartheta(u_{M_k,l}) - \vartheta(v_{M_k,l})\right| \\ \geq & \frac{1}{M_k}\sum^{l_k}_{l=l_k - \lceil  \epsilon_3 M_k \rceil}\left|\vartheta(u_{M_k,l}) - \vartheta(v_{M_k,l_k})\right|\\ \geq & \epsilon_3\epsilon_1>0
\end{align*}
for all $k\geq k_1$. This contradicts Lemma~\ref{lem: absolute uniform converg}. Thus we have proved that \eqref{eq:uniform convergence for continuous func} implies \eqref{eq:individual convergence for continuous func}. \cqfd

\section{Proof of Theorem \ref{thm:convergence rate band Toeplitz}}\label{prf:convergence rate band Toeplitz}

We first establish the following useful results.
\begin{lem}Let $u_0,u_1,\ldots,u_{N-1}\in \R$ be an unordered sequence of $N$ elements. We decreasingly arrange this sequence so that $u_{\rho(0)}\geq u_{\rho(1)}\geq \cdots \geq u_{\rho(N-1)}$. Then for any $r\in\{1,2\ldots,N-1\}$, we have
\begin{align*}
&\max_{1\leq r' \leq r}\max_{l\in[N-r'-1]}u_{\rho(l)} - u_{\rho(l+r')}\\ \leq & \max_{1\leq r'\leq r }~\max_{l\in[N-r'-1]}\left|u_{l} - u_{l+r'}\right|.
\end{align*}
\label{lem:error unordered and ordered sequence}\end{lem}
\noindent{\textbf{Proof}} (of Lemma \ref{lem:error unordered and ordered sequence}). The proof is straightforward for the case $r=1$. If the sequence is constant, then $$\max_{l\in[N-2]}u_{\rho(l)} - u_{\rho(l+1)}= \max_{l\in[N-2]}\left|u_{l} - u_{l+1}\right| = 0.$$ Suppose the sequence is not constant, i.e., there exist at least $l_1,l_2\in[N]$ so that $u_{l_1}\neq u_{l_2}$. Let  $$l' = \argmax_{l\in[N-2]}u_{\rho(l)} - u_{\rho(l+1)}$$
denote any point at which $u_{\rho(l)} - u_{\rho(l+1)}$ achieves its maximum.  Search the sequence $\left\{ u_l\right\}_{l\in[N]}$ to find $u_{l''}$ that is smaller than $u_{\rho(l')}$ and its index $l''$ is closest to $\rho(l')$. Thus $$\max\left\{\left|u_{l''} - u_{l''+1}\right|,\left|u_{l''} - u_{l''-1}\right| \right\}\geq \max_{l\in[N-2]}u_{\rho(l)} - u_{\rho(l+1)}.$$

Suppose $r\geq 2$. Similarly, the proof for a constant sequence is obvious. Suppose the sequence is not constant. Let
$$\left\{l',r'\right\} = \arg\max_{1\leq r'' \leq r}\max_{l\in[N-r''-1]}u_{\rho(l)} - u_{\rho(l+r'')}.$$ If there are several pairs $\left\{l',r' \right\}$ have the same values, we choose the one that $r'$ has the smallest value. If $r'=1$, the proof is similar to the case $r =1$. We suppose $r'\geq 2$. Thus there exist at least $r'$ elements that are smaller than $u_{\rho(l')}$ and only $r'-1$ elements that are greater than $u_{\rho(l'+r')}$ and smaller than $u_{\rho(l')}$.  Search the sequence $\left\{ u_l\right\}_{l\in[N]}$ to find $u_{l''}$ that is smaller than $u_{\rho(l')}$ and its index $l''$ is the $r'$-th closest to $\rho(l')$. Without loss of generality, suppose $l''< \rho(l')$.

If $u_{l''}\leq u_{\rho(l'+r')}$, we have
$$\max_{1\leq r''\leq r'}\left| u_{l''} - u_{l'' + r''}\right|  \geq u_{\rho(l')} - u_{\rho(l'+r')}$$
since there is at least one element in $\left\{u_l, l'' +1\leq l\leq l''+r'\right\}$ that is greater than or equal to $u_{\rho(l')}$.

If $u_{l''}> u_{\rho(l'+r')}$, there exists $l''  \leq l'''\leq 2\rho(l') - l''$ such that $u_{l'''}$ is smaller than or equal to $u_{\rho(l'+r')}$ (otherwise, there are $r'$ elements that are greater than $u_{\rho(l'+r')}$ and smaller than $u_{\rho(l')}$). Also near $u_{l'''}$, there must exist at least one element that is not smaller than $u_{\rho(l')}$. Then
\begin{align*}
&\max_{1\leq r''\leq r'}\max\left\{\left|u_{l'''} - u_{l'''+r''}  \right|,\left|u_{l'''} - u_{l'''-r''}  \right| \right\}\\ \geq & u_{\rho(l')} - u_{\rho(l'+r')}.
\end{align*}
This completes the proof. \cqfd

In words, the largest error between the contiguous elements of a sequence is not magnified when the sequence is rearranged in decreasing (or increasing) order.

The following result establishes that the largest error between two sequences is not magnified when both of the sequences are  rearranged in decreasing (or increasing) order.

\begin{lem}Let $u_0,\ldots,u_{N-1}\in \R$ and $v_0,\ldots,v_{N-1}\in \R$ be two unordered sequences of $N$ elements. We decreasingly arrange these sequences so that $u_{\rho(0)}\geq u_{\rho(1)}\geq \cdots u_{\rho(N-1)}$ and $v_{\rho(0)}\geq v_{\rho(1)}\geq \cdots v_{\rho(N-1)}$. Then
\begin{align*}
\max_{l\in[N-1]}\left|u_{\rho(l)} - v_{\rho(l)}\right|\leq \max_{l\in[N-1]}\left|u_{l} - v_{l}\right|.
\end{align*}
\label{lem:error unordered and ordered sequence 2}\end{lem}
\noindent{\textbf{Proof}} (of Lemma \ref{lem:error unordered and ordered sequence 2}). Let $$r' = \argmax_{r\in[N-1]}\left|u_{\rho(r)} - v_{\rho(r)}\right|$$
denote any point at which $\left|u_{\rho(r)} - v_{\rho(r)}\right|$ achieves its maximum and let $l'$ be the index of $u_{\rho(r')}$. Without loss of generality, we suppose $u_{\rho(r')}\geq v_{\rho(r')}$. If $v_{l'}\leq v_{\rho(r')}$, we have $u_{l'} - v_{l'}\geq u_{\rho(r')}- v_{\rho(r')}$. Otherwise suppose $v_{l'}> v_{\rho(r')}$, which implies $r'\geq 1$. Since there are only $r'$ elements in $\left\{ u_l\right\}_{l\in [N]}$ that are greater than $u_{\rho(r')}$ and $r'$ elements in $\left\{ v_l\right\}_{l\in [N]}$ that are greater than $v_{\rho(r')}$, there must exist $l''$ such that $u_{l''}\geq u_{\rho(r')}$ and $v_{l''}\leq v_{\rho(r')}$. Hence $$u_{l''} - v_{l''}\geq u_{\rho(r')}- v_{\rho(r')}.$$ \cqfd

\begin{lem}\cite[Sturmian separation theorem]{Horn1985MatrixAnalysis}
Let $\bm A_N$ be an $N\times N$ Hermitian matrix and let $[\bm A_N]_{N-1}$ be the $\left(N-1\right) \times \left( N-1\right)$ matrix obtained by deleting the last column and the last row of $\bm A_N$. Also let $\lambda_0(\bm A_N)\geq \cdots\geq\lambda_{N-1}(\bm A_N)$ and $\lambda_0([\bm A_N]_{N-1})\geq \cdots\geq\lambda_{N-2}([\bm A_N]_{N-1})$ respectively denote the descending eigenvalues of $\bm A_N$ and $[\bm A_N]_{N-1}$. Then
\begin{align*}
\lambda_l(\bm A_N)\geq \lambda_l([\bm A_N]_{N-1})\geq \lambda_{l+1}(\bm A_N)
\end{align*}
for all $0\leq l \leq N-2$.
\end{lem}
The above Sturmian separation theorem forms the foundation of the following analysis. We note that Zizler et.al.~\cite{zizler2002finer} utilized the Sturmian separation theorem to prove a refinement of Szeg\H{o}'s asymptotic formula in terms of the number of eigenvalues inside a given interval.

\vspace{.1in}
Now we are well equipped to prove Theorem \ref{thm:convergence rate band Toeplitz}. In what follows, we consider $N> 2r$. Note that in this case $\widetilde{\bm C}_N$ is equivalent to $\widehat{\bm C}_N$ and the eigenvalues of $\widetilde{\bm C}_N$ are the DFT samples of $S_{N-1}(f) = \widetilde h(f) =\sum_{k = -r}^{r} h[k]e^{j2\pi fk}$. Recall that $[\bm H_N]_{N-r}$  is the $\left(N-r\right)\times \left(N-r\right)$ matrix obtained by deleting the last $r$ columns and the last $r$ rows of $\bm H_N$. Similar notation holds for $[\widetilde{\bm C}_N]_{N-r}$.

Note that $[\bm H]_{N-r}$ is exactly the same as $[\widetilde{\bm C}_N]_{N-r}$ as they have the same elements when $N> 2r$. Thus $[\bm H]_{N-r}$ and $[\widetilde{\bm C}_N]_{N-r}$ have the same eigenvalues. Let $\lambda_l([\widetilde{\bm C}_{N}]_{N-r})$  be permuted such that $$\lambda_{\rho(0)}([\widetilde{\bm C}_{N}]_{N-r}) \geq\cdots \geq \lambda_{\rho(N-r-1)}([\widetilde{\bm C}_{N}]_{N-r}).$$

We first consider the simple case when $r=1$. It follows from the Sturmian separation theorem that
\begin{align*}
&\lambda_l(\bm H_N) \geq \lambda_l([\bm H_N]_{N-1} )\geq \lambda_{l+1}(\bm H_N),\\
& \lambda_{\rho(l)}(\widetilde{\bm C}_N) \geq \lambda_{\rho(l)}([\widetilde{\bm C}_N]_{N-1} )\geq \lambda_{\rho(l+1)}(\widetilde{\bm C}_N)
\end{align*}
for all $0\leq l\leq N-2$. This implies the following relationship between $\lambda_l(\bm H_N)$ and $\lambda_{\rho(l)}(\widetilde{\bm C}_N)$
\begin{equation}\label{eq:nest relation betw eig H and eig C}
\begin{split}
\lambda_l(\bm H_N)\leq & \lambda_{l-1}([\bm H_N]_{N-1} ) = \lambda_{\rho(l-1)}([\widetilde{\bm C}_N]_{N-1} )\\ \leq & \lambda_{\rho(l-1)}(\widetilde{\bm C}_N), ~\forall ~ l = 1,2,\ldots,N-1,\\
\lambda_l(\bm H_N)\geq & \lambda_{l}([\bm H_N]_{N-1} ) = \lambda_{\rho(l)}([\widetilde{\bm C}_N]_{N-1} ) \\ \geq & \lambda_{\rho(l+1)}(\widetilde{\bm C}_N), ~\forall ~ l = 0,1,\ldots,N-2
\end{split}\end{equation}
which give
\begin{align*}
&\left|\lambda_l\left(\bm H_N\right) - \lambda_{\rho(l)}\left( \widetilde{\bm C}_N\right)\right| \\
= &\max\left\{ \lambda_l\left(\bm H_N\right) - \lambda_{\rho(l)}\left( \widetilde{\bm C}_N\right), \right. \\& \quad \quad \quad \left. \lambda_{\rho(l)}\left( \widetilde{\bm C}_N\right)-\lambda_l\left(\bm H_N\right)  \right\}\\
\leq & \max\left\{ \lambda_{\rho(l-1)}\left( \widetilde{\bm C}_N\right) - \lambda_{\rho(l)}\left( \widetilde{\bm C}_N\right),\right.\\ & \quad\quad\quad \left. \lambda_{\rho(l)}\left( \widetilde{\bm C}_N\right) -\lambda_{\rho(l+1)}\left( \widetilde{\bm C}_N\right)  \right\}
\end{align*}
for all $1\leq l\leq N-2.$ Applying Lemma~\ref{lem:error unordered and ordered sequence} with $r=1$, we obtain
\begin{align*}
&\max_{1\leq l\leq N-2} \left|\lambda_l\left(\bm H_N\right) - \lambda_{\rho(l)}\left( \widetilde{\bm C}_N\right)\right|\\ \leq & \max_{0\leq l\leq N-2} \lambda_{\rho(l)}\left( \widetilde{\bm C}_N\right) -\lambda_{\rho(l+1)}\left( \widetilde{\bm C}_N\right)\\ \leq & \max_{0\leq l\leq N-2} \left|\lambda_{l}\left( \widetilde{\bm C}_N\right) -\lambda_{l+1}\left( \widetilde{\bm C}_N\right)\right|.
\end{align*}
Note that $\widetilde h(f) $ is Lipschitz continuous since it is continuously differentiable. There exists a Lipschitz constant $K$ such that, for all $f_1$ and $f_2$ in $[0,1]$,
\begin{align*}
\left|\widetilde h(f_1) - \widetilde h(f_2) \right|\leq K \left| f_1 - f_2\right|.
\end{align*}
From the fact that the eigenvalues of $\widetilde{\bm C}_N$ are the DFT samples of $\widetilde h(f)$, i.e., $\lambda_{l}\left( \widetilde{\bm C}_N\right) = \widetilde h(\frac{l}{N})$, it follows that
\begin{equation}\begin{split}
&\max_{1\leq l\leq N-2} \left|\lambda_l\left(\bm H_N\right) - \lambda_{\rho(l)}\left( \widetilde{\bm C}_N\right)\right|\\ \leq & \max_{0\leq l\leq N-2} \left|\lambda_{l}\left( \widetilde{\bm C}_N\right) -\lambda_{l+1}\left( \widetilde{\bm C}_N\right)\right|\\
\leq & \max_{0\leq l\leq N-2} \left|\widetilde h(\frac{l}{N}) -\widetilde h(\frac{l+1}{N})\right| \leq K \frac{1}{N}.
\end{split}\label{eq:prf fir convergence 1}\end{equation}
Utilizing the fact that $\lambda_{0}(\bm H_N)\leq \max_{f\in[0,1]}\widetilde h(f)$ and $\lambda_{N-1}(\bm H_N)\geq \min_{f\in[0,1]}\widetilde h(f)$ (see Lemma~\ref{lem:bounds Pearl's eigenvalues}) and applying \eqref{eq:nest relation betw eig H and eig C} with $l=0$ which gives
\[
\lambda_0\left(\bm H_N\right) \geq \lambda_{\rho(1)}\left( \widetilde{\bm C}_N\right),
\] we have
\begin{align*}
&\left|\lambda_0\left(\bm H_N\right) - \lambda_{\rho(0)}\left( \widetilde{\bm C}_N\right)\right|\\
= &\max\left\{ \lambda_0\left(\bm H_N\right) - \lambda_{\rho(0)}\left( \widetilde{\bm C}_N\right), \right. \\& \quad \quad \quad \left. \lambda_{\rho(0)}\left( \widetilde{\bm C}_N\right)-\lambda_0\left(\bm H_N\right)  \right\}\\
 \leq & \max\left\{ \max_{f\in[0,1]}\widetilde h(f) -\lambda_{\rho(0)}\left( \widetilde{\bm C}_N\right), \right. \\& \quad \quad \quad \left. \lambda_{\rho(0)}\left( \widetilde{\bm C}_N\right) - \lambda_{\rho(1)}\left( \widetilde{\bm C}_N\right)  \right\}\\ \leq & K\frac{1}{N},
\end{align*}
 where the second inequality follows because $\lambda_l\left( \widetilde{\bm C}_N\right)$ are uniform samples of $\widetilde h(f)$ with grid size $\frac{1}{N}$. Similarly, we have
\begin{align*}
&\left|\lambda_{N-1}\left(\bm H_N\right) - \lambda_{\rho(N-1)}\left( \widetilde{\bm C}_N\right)\right|\\
= &\max\left\{\lambda_{\rho(N-1)}\left( \widetilde{\bm C}_N\right) - \lambda_{N-1}\left(\bm H_N\right), \right. \\& \quad \quad \quad \left. \lambda_{N-1}\left(\bm H_N\right) - \lambda_{\rho(N-1)}\left( \widetilde{\bm C}_N\right)  \right\}\\
\leq & \max\left\{ \lambda_{\rho(N-1)}\left( \widetilde{\bm C}_N\right) - \min_{f\in[0,1]}\widetilde h(f) ,\right. \\ & \quad \quad \quad \left.  \lambda_{\rho(N-2)}\left( \widetilde{\bm C}_N\right) - \lambda_{\rho(N-1)}\left( \widetilde{\bm C}_N\right)  \right\}\\ \leq & K\frac{1}{N}.
\end{align*}
Along with \eqref{eq:prf fir convergence 1}, we conclude
\begin{align*}
\max_{0\leq l\leq N-1} \left|\lambda_l\left(\bm H_N\right) - \lambda_{\rho(l)}\left( \widetilde{\bm C}_N\right)\right|\leq  K \frac{1}{N}.
\end{align*}

Now we consider the case $r>1$. Repeatedly applying the Sturmian separation theorem $r$ times yields
\begin{align*}
&\lambda_l(\bm H_N) \geq \lambda_l([\bm H_N]_{N-r} )\geq \lambda_{l+r}(\bm H_N),\\
& \lambda_{\rho(l)}(\widetilde{\bm C}_N) \geq \lambda_{\rho(l)}([\widetilde{\bm C}_N]_{N-r} )\geq \lambda_{\rho(l+r)}(\widetilde{\bm C}_N)
\end{align*}
for all $0\leq l\leq N-r-1$. Noting that $[\bm H_N]_{N-r}$ is the same as $[\widetilde{\bm C}_N]_{N-r}$, we have
\begin{equation*}
\begin{split}
\lambda_l(\bm H_N)\leq & \lambda_{l-r}([\bm H_N]_{N-r} ) = \lambda_{\rho(l-r)}([\widetilde{\bm C}_N]_{N-r} )\\ \leq & \lambda_{\rho(l-r)}(\widetilde{\bm C}_N), ~\forall ~ l = r,r+1,\ldots,N-1,\\
\lambda_l(\bm H_N)\geq & \lambda_{l}([\bm H_N]_{N-r} ) = \lambda_{\rho(l)}([\widetilde{\bm C}_N]_{N-r} ) \\ \geq & \lambda_{\rho(l+r)}(\widetilde{\bm C}_N), ~\forall ~ l = 0,1,\ldots,N-r-1
\end{split}\end{equation*}
which give
\begin{align*}
&\max_{r\leq l\leq N-r-1} \left|\lambda_l\left(\bm H_N\right) - \lambda_{\rho(l)}\left( \widetilde{\bm C}_N\right)\right|\\
= &\max_{r\leq l\leq N-r-1} \max\left\{ \lambda_l\left(\bm H_N\right) - \lambda_{\rho(l)}\left( \widetilde{\bm C}_N\right), \right. \\& \quad \quad \quad \quad \quad \quad \quad \quad \left. \lambda_{\rho(l)}\left( \widetilde{\bm C}_N\right)-\lambda_l\left(\bm H_N\right)  \right\}\\
\leq &\max_{r\leq l\leq N-r-1} \max\left\{ \lambda_{\rho(l-r)}\left( \widetilde{\bm C}_N\right) - \lambda_{\rho(l)}\left( \widetilde{\bm C}_N\right), \right. \\& \quad \quad \quad \quad \quad \quad \quad \quad \left. \lambda_{\rho(l)}\left( \widetilde{\bm C}_N\right)-\lambda_{\rho(l+r)}\left( \widetilde{\bm C}_N\right)  \right\}\\
\leq & \max_{0\leq l\leq N-r-1} \lambda_{\rho(l)}\left( \widetilde{\bm C}_N\right) -\lambda_{\rho(l+r)}\left( \widetilde{\bm C}_N\right) \\
\leq & \max_{1\leq r'\leq r}~\max_{0\leq l\leq N-r'-1} \left|\lambda_{l}\left( \widetilde{\bm C}_N\right) -\lambda_{l+r'}\left( \widetilde{\bm C}_N\right)\right|\\
 \leq  & K\frac{r}{N},
\end{align*}
where the third inequality follows from Lemma \ref{lem:error unordered and ordered sequence}.
Since $$\lambda_{r-1}(\bm H_N)\leq\cdots\leq\lambda_{0}(\bm H_N)\leq \max_{f\in[0,1]}\widetilde h(f),$$
we bound $\left|\lambda_{r'}\left(\bm H_N\right) - \lambda_{\rho(r')}\left( \widetilde{\bm C}_N\right)\right|, r'\leq r-1$ by considering the following two cases: if $\lambda_{\rho(r')}\left( \widetilde{\bm C}_N\right) \leq \lambda_{r'}\left(\bm H_N\right)$, we have
\begin{align*}&\lambda_{r'}\left(\bm H_N\right) - \lambda_{\rho(r')}\left( \widetilde{\bm C}_N\right)\\
\leq & \max_{f\in[0,1]}\widetilde h(f) - \lambda_{\rho(r-1)}\left( \widetilde{\bm C}_N\right)\leq K\frac{r}{N};
\end{align*}
if $\lambda_{\rho(r')}\left( \widetilde{\bm C}_N\right) > \lambda_{r'}\left(\bm H_N\right)$, we have
\begin{align*}
&\lambda_{\rho(r')}\left( \widetilde{\bm C}_N\right) - \lambda_{r'}\left(\bm H_N\right)\\\leq & \lambda_{\rho(r')}\left( \widetilde{\bm C}_N\right) - \lambda_{\rho(r'+r)}\left( \widetilde{\bm C}_N\right)\\
\leq & \max_{1\leq r''\leq r}\max_{0\leq l\leq N-r''-1} \left|\lambda_{l}\left( \widetilde{\bm C}_N\right) -\lambda_{l+r''}\left( \widetilde{\bm C}_N\right)\right|\\
\leq & K\frac{r}{N}
\end{align*}
where the second line follows because
$\lambda_{r'}\left(\bm H_N\right) \geq \lambda_{\rho(r'+r)}\left( \widetilde{\bm C}_N\right)$ and the third line follows from Lemma~\ref{lem:error unordered and ordered sequence}.
Thus we have
\begin{align*}
&\left|\lambda_{r'}\left(\bm H_N\right) - \lambda_{\rho(r')}\left( \widetilde{\bm C}_N\right)\right| \leq K\frac{r}{N}
\end{align*}
for all $0\leq r'\leq r-1$. Similarly,
\begin{align*}
&\left|\lambda_{N-r'}\left(\bm H_N\right) - \lambda_{\rho(N-r')}\left( \widetilde{\bm C}_N\right)\right| \\ \leq & \max\left\{ \lambda_{\rho(N-r')}\left( \widetilde{\bm C}_N\right) - \min_{f\in[0,1]}\widetilde h(f), \right. \\ &\quad\quad\quad \left. \lambda_{\rho(N-r'-r)}\left( \widetilde{\bm C}_N\right) - \lambda_{\rho(N-r')}\left( \widetilde{\bm C}_N\right)  \right\}\\
\leq & \max\left\{K\frac{r}{N},K\frac{r}{N} \right\}= K\frac{r}{N},
\end{align*}
for all $1\leq r'\leq r$. Therefore,
\begin{align*}
\max_{0\leq l\leq N-1} \left|\lambda_l\left(\bm H_N\right) - \lambda_{\rho(l)}\left( \widetilde{\bm C}_N\right)\right|\leq  Kr \frac{1}{N}.
\end{align*}
for all $N> 2r$.

Note that $S_{r+1}(f) = S_{r+2}(f) = \cdots = S_{N-1}(f)$ which gives $$\sigma_N(f) = \frac{\sum_{n=0}^{N-1}S_n(f)}{N} = \frac{\sum_{n=0}^{r}S_n(f)}{N}+ \frac{N-r-1}{N}S_{r+1}(f).$$ Thus
\begin{align*}
\left|\sigma_N(f) - S_{N-1}(f)\right| =& \left|\frac{\sum_{n=0}^{r}S_n(f)}{N} - \frac{r+1}{N}S_{r+1}(f)\right|\\ =& \left|\sum_{n=0}^{r}S_n(f) - (r+1)S_{r+1}(f) \right|\frac{1}{N}\\
 = & O(\frac{1}{N})
\end{align*}
uniformly on $[0,1]$ as $N\rightarrow \infty$. Therefore,
\begin{align*}
&\max_{0\leq l\leq N-1}\left|\lambda_l(\overline{\bm C}_N) - \lambda_l(\widetilde{\bm C}_N)\right|\\ =& \max_{0\leq l\leq N-1}\left|\sigma_N(\frac{l}{N}) - S_{N-1}(\frac{l}{N})\right| = O(\frac{1}{N})
\end{align*}
as $N \rightarrow \infty$. Finally,
\begin{align*}
&\max_{0\leq l\leq N-1}\left|\lambda_{\rho(l)}(\overline{\bm C}_N) - \lambda_l(\bm H_N)\right| \\ =& \max_{0\leq l\leq N-1}\left|\lambda_{\rho(l)}(\overline{\bm C}_N) - \lambda_{\rho(l)}(\widetilde{\bm C}_N) + \lambda_{\rho(l)}(\widetilde{\bm C}_N) -\lambda_l(\bm H_N)\right|\\
 \leq & \max_{0\leq l\leq N-1} \left|\lambda_{\rho(l)}(\overline{\bm C}_N) - \lambda_{\rho(l)}(\widetilde{\bm C}_N) \right|\\ & +\max_{0\leq l\leq N-1} \left| \lambda_{\rho(l)}(\widetilde{\bm C}_N) -\lambda_l(\bm H_N)\right|\\
 \leq &\max_{0\leq l\leq N-1}\left|\lambda_l(\overline{\bm C}_N) - \lambda_l(\widetilde{\bm C}_N)\right| \\ &+ \max_{0\leq l\leq N-1} \left| \lambda_{\rho(l)}(\widetilde{\bm C}_N) -\lambda_l(\bm H_N)\right|\\ = & O(\frac{1}{N})
\end{align*}
as $N\rightarrow \infty$, where the second inequality follows from Lemma~\ref{lem:error unordered and ordered sequence 2}. \cqfd

\section{Proof of Theorem \ref{thm:uniform to individual}}\label{prf:uniform to individual}

\begin{thm}(Riemann-Lebesgue theorem~\cite[Theorem 7.48]{apostol1974mathematical})
The function $g(x)\in L^{\infty}\left( [a,b]\right)$ is Riemann integrable over $[a,b]$ if and only if it is continuous almost everywhere in $[a,b]$.
\end{thm}

Despite the fact that $F_g(\alpha)$ is right continuous and non-decreasing everywhere, some stronger results about $F_g(\alpha)$ can be obtained at some point $\alpha$ since $g(x)$ is Riemann integrable.
\begin{lem}
Suppose $g(x):[c,d]\rightarrow \bm [a,b]$ is Riemann integrable and let $F_g(\alpha)$ be defined as in~\eqref{eq:def F_g}. Then $F_g(\alpha)$ is strictly increasing at $\alpha$ if $\alpha\in\textup{int}\left(\ess\mathcal{R}(g)\right)$, i.e., there exists $\epsilon>0$ such that, for every pair $(\alpha_1,\alpha_2)$ such that $\alpha -\epsilon<\alpha_1<\alpha<\alpha_2<\alpha+\epsilon$, $F_g(\alpha_1)<F_g(\alpha)<F_g(\alpha_2)$.
\label{lem:F stricitly increasing}\end{lem}
\vspace{0.3cm}
\noindent{\textbf{Proof}} (of Lemma \ref{lem:F stricitly increasing}). Since $g(x):[c,d]\rightarrow \bm [a,b]$ is Riemann integrable and $\alpha\in\textup{int}\left(\ess\mathcal{R}(g)\right)$, there exists $\epsilon$ such that $\left(\alpha - \epsilon, \alpha + \epsilon \right) \subset \ess\mathcal{R}(g)$. Let $\alpha_1$ be an arbitrary value such that $\alpha -\epsilon<\alpha_1<\alpha$ and let $\alpha_1' = \frac{\alpha + \alpha_1}{2} \in \ess\mathcal{R}(g)$. It follows from the definition of essential range that
$$\mu\left\{x\in[c,d]: \left|g(x) - \alpha_1'\right|<\frac{\alpha-\alpha_1}{2}  \right\} >0 .$$
Thus, we obtain
\begin{align*}F_g(\alpha) - F_g(\alpha_1) =& \frac{1}{d-c}\mu\left\{ x\in[c,d]: \alpha_1<g(x)\leq \alpha  \right\}\\ \geq &\frac{1}{d-c}\mu\left\{ x\in[c,d]: \alpha_1<g(x)< \alpha  \right\}>0.
\end{align*}
Similarly, we have $F_g(\alpha)< F_g(\alpha_2)$ for $\alpha<\alpha_2<\alpha+\epsilon$.  \cqfd
%

We are now ready to prove the main part using the same approach that was used to prove Theorem~\ref{thm:uniform to individual for contunuous fuc}.

First, we show that \eqref{eq:individual convergence} implies \eqref{eq:uniform convergence}. This part is the same as those in Appendix~\ref{prf:uniform to individual for contunuous fuc}.

Now let us show that \eqref{eq:uniform convergence} implies \eqref{eq:individual convergence}. We prove the statement \eqref{eq:uniform convergence} $\Rightarrow$ \eqref{eq:individual convergence} by contradiction. Suppose \eqref{eq:individual convergence} is not true, i.e., there exists an increasing sequence $\left\{M_{k'}\right\}_{k'=1}^{\infty}$  and $\epsilon_1>0$ such that
\[
\max_{l\in[M_{k'}]}\left| u_{M_{k'},l} - v_{M_{k'},l} \right|\geq 2\epsilon_1
\]
for all $k'\geq 1$. Let $l_{k'} = \arg\max_{l\in[M_{k'}]}\left| u_{M_{k'},l} - v_{M_{k'},l} \right|$ denote any point at which $\left| u_{M_{k'},l} - v_{M_{k'},l} \right|$ achieves its maximum, which implies  $\left| u_{M_{k'},l_{k'}} - v_{M_{k'},l_{k'}} \right|\geq 2\epsilon_1$. Without loss of generality, we suppose $\{M_{k},l_{k}\}_{k=1}^\infty$ is a subsequence of $\{M_{k'},l_{k'}\}_{k'=1}^\infty$ such that
$u_{M_{k},l_{k}}\leq  v_{M_{k},l_{k}}$, i.e., $u_{M_{k},l_{k}}\leq  v_{M_{k},l_{k}} - 2\epsilon_1$. We also suppose $F_g$ is continuous at $u_{M_k,l_k}$ because otherwise, one can always pick a $\widehat u_{M_k,l_k}\in \mbox{int}\left(\ess\mathcal{R}(g)\right)$ that is close enough to $u_{M_k,l_k}$ and such that $F_g$ is continuous at $\widehat u_{M_k,l_k}$ since $F_g$ is continuous almost everywhere.

By assumption, $u_{M_k,l_k}\in \mbox{int}\left(\ess\mathcal{R}(g)\right)$. Noting that $F_g$ is right continuous everywhere and strictly increasing at all $u_{M_k,l_k}$ (which is shown in Lemma~\ref{lem:F stricitly increasing}), there exist $\epsilon_2(u_{M_k,l_k})>0$ (which depends on $u_{M_k,l_k}$ and for notational simplicity, we rewrite as $\epsilon_2$)  and $\epsilon_3>0$  (which is independent of $u_{M_k,l_k}$) such that $\epsilon_2\leq \frac{\epsilon_1}{2}$, $F_g$ is continuous at $u_{M_k,l_k} + \epsilon_2$, and
\begin{align}
F_g(u_{M_k,l_k} + \epsilon_2) \geq F_g(u_{M_k,l_k}) + 3\epsilon_3.
\label{eq:F_u strictly continuous}\end{align}
 Lemma~\ref{lem:convergence cdf} indicates that
$$\lim_{M_k\rightarrow \infty}F_{u_{M_k}}(\alpha) = F_g(\alpha)$$
for every point $\alpha$ at which $F_g$ is continuous.
Thus there exists $k_0\in \N$ such that
\begin{equation}\begin{split}
\left| F_{u_{M_k}}\left(u_{M_k,l_k} \right) - F_g\left(u_{M_k,l_k} \right)\right| \leq & \epsilon_3, \\ \left| F_{u_{M_k}}\left(u_{M_k,l_k} + \epsilon_2 \right) - F_g\left(u_{M_k,l_k} + \epsilon_2 \right)\right| \leq & \epsilon_3
\end{split}
\label{eq:cdf equivalent}\end{equation}
for all $k\geq k_0$.
Thus, we have
\begin{align*}
&F_{u_{M_k}}(u_{M_k,l_k} + \epsilon_2) -F_{u_{M_k}}(u_{M_k,l_k})
\\ = &  \left( F_{u_{M_k}}(u_{M_k,l_k} + \epsilon_2)  - F_{g}(u_{M_k,l_k} + \epsilon_2) \right) \\
& +  \left(F_{g}(u_{M_k,l_k} + \epsilon_2 ) - F_{g}(u_{M_k,l_k} )\right)\\ & +  \left( F_{g}(u_{M_k,l_k})  - F_{u_{M_k}}(u_{M_k,l_k}) \right) \\ \geq &  F_{g}(u_{M_k,l_k} + \epsilon_2 )  - F_{g}(u_{M_k,l_k}) \\ &- \left| F_{u_{M_k}}(u_{M_k,l_k} + \epsilon_2)  - F_{g}(u_{M_k,l_k} + \epsilon_2) \right|  \\ & -\left| F_{g}(u_{M_k,l_k})  - F_{u_{M_k}}(u_{M_k,l_k})\right|\\
 \geq & 3\epsilon_3 -\epsilon_3 -\epsilon_3 = \epsilon_3
\end{align*}
for all $k\geq k_0$, where the last line follows from \eqref{eq:F_u strictly continuous} and \eqref{eq:cdf equivalent}.
Note that the above equation is equivalent to
$$\frac{1}{M_k}\#\left\{l\in[M_k], u_{M_k,l_k} < u_{M_k,l}\leq u_{M_k,l_k} + \epsilon_2  \right\}\geq \epsilon_3.$$
Then
$$0\leq u_{M_k,l_k - \lceil  \epsilon_3 M_k \rceil} - u_{M_k, l_k} \leq u_{M_k,l_k} + \epsilon_2 - u_{M_k,l_k} =  \epsilon_2,$$
which implies
\begin{align*}&v_{M_k, l_k } - u_{M_k,l_k - \lceil  \epsilon_3 M_k \rceil}\\ = & v_{M_k, l_k } - u_{M_k, l_k } + u_{M_k, l_k } - u_{M_k,l_k - \lceil  \epsilon_3 M_k \rceil}\\
 \geq & 2\epsilon_1 -\epsilon_2\geq 2\epsilon_1 - \epsilon_1 \geq \epsilon_1.
\end{align*}
Now taking $\vartheta(t) = t$, we obtain
\begin{align*}&\frac{1}{M_k}\sum_{l=0}^{M_k-1}\left|\vartheta(u_{M_k,l}) - \vartheta(v_{M_k,l})\right| \\ \geq &\frac{1}{M_k}\sum^{l_k}_{l=l_k - \lceil  \epsilon_3 M_k \rceil}\left|\vartheta(u_{M_k,l}) - \vartheta(v_{M_k,l})\right| \\ \geq & \frac{1}{M_k}\sum^{l_k}_{l=l_k - \lceil  \epsilon_3 M_k \rceil}\left|\vartheta(u_{M_k,l}) - \vartheta(v_{M_k,l_k})\right|\\ \geq & \epsilon_3\epsilon_1>0
\end{align*}
for all $k\geq k_1$. This contradicts Lemma~\ref{lem: absolute uniform converg}. \cqfd

\section{Proof of Theorem \ref{thm:bounds extreme eigenvalues}}\label{prf:bounds extreme eigenvalues}
\begin{lem}
Let $D_N(f): = \frac{\sin\left(\pi N f  \right)}{\sin\left(\pi f \right)}$ denote the Dirichlet kernel. Fix $0<W<\frac{1}{2}$. We have
\begin{align*}
\int_{0}^{1}\left|D_N(f)\right|^2 df & = N, \enskip \forall \enskip N\in \N,\\
\int_{W}^{1-W} \left|D_N(f)\right|^2 df & = O(1), \enskip \mbox{when}\enskip  N\rightarrow \infty.
\end{align*}
\label{lem: dirichlet kernel}\end{lem}

\vspace{0.3cm}
\noindent{\textbf{Proof}} (of Lemma \ref{lem: dirichlet kernel}). Noting that $D_N(f) = \frac{\sin\left(\pi N f  \right)}{\sin\left(\pi f \right)} = \frac{e^{j2\pi f N}}{e^{j\pi f}}\sum_{n=0}^{N-1}e^{j2\pi f n}$, we have
\begin{align*}
\left|D_N(f)\right|^2 =& \left| \sum_{n=0}^{N-1}e^{j2\pi f n} \right|^2 = \left( \sum_{n=0}^{N-1}e^{j2\pi f n} \right) \left( \sum_{m=0}^{N-1}e^{-j2\pi f m} \right)\\ =& \sum_{n=0}^{N-1}\sum_{m=0}^{N-1}e^{j2\pi f(n - m)}.
\end{align*}
It follows that
\begin{align*}
\int_0^1 \left|D_N(f)\right|^2 df =& \int_0^1 \sum_{n=0}^{N-1}\sum_{m=0}^{N-1}e^{j2\pi f(n - m)} df \\ =& \sum_{n=0}^{N-1}\sum_{m=0}^{N-1} \int_0^1e^{j2\pi f(n - m)} df = N.
\end{align*}
Fix $0<W<\frac{1}{2}$. For any $f\in [W,1-W]$, $\left|D_N(f)\right|$ is bounded above by $\frac{1}{\sin\left(\pi W \right)}$. Therefore,
\begin{align*}
\int_{W}^{1-W} \left|D_N(f)\right|^2 df \leq \int_{W}^{1-W}\frac{1}{\sin^2\left(\pi W \right)}df \leq \frac{1}{\sin^2\left(\pi W \right)}.
\end{align*}
\cqfd

Since $\widetilde h$ is bounded and Riemann integrable over $[0,1]$, it follows from the Riemann-Lebesgue theorem that $\widetilde h$ is continuous almost everywhere in $[0,1]$. Thus we can select $f_0\in[0,1]$ and a positive number $W$ such that
$$\left|\widetilde h(f) - \esssup \widetilde h\right|\leq \frac{\epsilon}{4}$$
holds almost everywhere for $\left| f - f_0 \right|\leq W$. For any $\bm v\in \mathbb{C}^N$, we have
\begin{equation}\begin{split}
&\langle \bm H_N \bm v, \bm v \rangle \\  = & \int_{0}^{1}\left|\widetilde{\bm v}(f)\right|^2\widetilde h(f)df\\
= & \int_{f_0 - W}^{f_0 + W}\left|\widetilde{\bm v}(f)\right|^2\widetilde h(f)df +  \int_{\scriptstyle f \in [0,1] \atop \scriptstyle f\notin [f_0 - W,f_0 +W]} \left|\widetilde{\bm v}(f)\right|^2\widetilde h(f)df \\
\geq & \left(\esssup\widetilde h - \frac{\epsilon}{4} \right)\int_{f_0 - W}^{f_0 + W}\left|\widetilde{\bm v}(f)\right|^2\widetilde df \\ &- \max\left( \left|\essinf\widetilde h\right|, \left|\esssup\widetilde h\right| \right)\cdot\int_{\scriptstyle f \in [0,1] \atop \scriptstyle f\notin [f_0 - W,f_0 +W]} \left|\widetilde{\bm v}(f)\right|^2\widetilde df.
\label{eq:eig bound}
\end{split}\end{equation}


The DTFT  of $\bm e_{l/N}$ is
\begin{align}
\widetilde {\bm e}_{l/N}(f) = \frac{e^{-j\pi N\left(f - \frac{l}{\sqrt N} \right)}}{e^{-j\pi \left(f - \frac{l}{N} \right)}}D_N(f - \frac{l}{N}).
\label{eq:DTFT e}\end{align}
Fix $\widetilde h$, $\epsilon$ and $W$. If $N\geq \frac{1}{W}$, there always exists $l'$ such that $\left| \frac{l'}{N} - f_0\right| \leq \frac{W}{2}$. It follows from Lemma~\ref{lem: dirichlet kernel} that
\begin{align*}
\int_{\frac{l'}{N} - \frac{W}{2}}^{\frac{l'}{N} + \frac{W}{2}} \left|\frac{1}{\sqrt{N}}\widetilde {\bm e}_{l'/N}(f)\right|^2df &= 1 - \frac{1}{N}\int_{\frac{W}{2}}^{1 - \frac{W}{2}}\left|D_N(f)\right|^2df \\ &= 1-o(1)
\end{align*}
as $N\rightarrow \infty$. Note that $[\frac{l'}{N} - \frac{W}{2}, \frac{l'}{N} + \frac{W}{2}]\subset [f_0 - W, f_0 + W]$. Thus there exists $N_1\in\N$ such that for all $N\geq \max\left\{N_1,\frac{1}{W}\right\}$
\begin{equation}\begin{split}
& \int_{f_0 - W}^{f_0 + W}\left|\frac{1}{\sqrt{N}}\widetilde {\bm e}_{l'/N}(f)\right|^2\widetilde df \geq 1-\frac{\epsilon}{4\left|\esssup\widetilde h\right|},\\
& \int_{\scriptstyle f \in [0,1] \atop \scriptstyle f\notin [f_0 - W,f_0 +W]} \left|\frac{1}{\sqrt{N}}\widetilde {\bm e}_{l'/N}(f)\right|^2\widetilde df \\
& \quad\quad \leq  \frac{\epsilon}{2\cdot \max\left( \left|\essinf\widetilde h\right|, \left|\esssup\widetilde h\right| \right)}.
\end{split}
\label{eq: dpss bound 2}\end{equation}
Combining \eqref{eq:eig bound} and \eqref{eq: dpss bound 2} yields \begin{align*}
\lambda_{l'}\left(\overline{\bm C}_N \right) = & \langle \bm H_N \frac{1}{\sqrt N}\bm e_{l'/N},
\frac{1}{\sqrt N}\bm e_{l'/N} \rangle \\
\geq & \left(\esssup\widetilde h - \frac{\epsilon}{4} \right)\left(1 - \frac{\epsilon}{4\left|\esssup\widetilde h\right|} \right) - \frac{\epsilon}{2} \\
 \geq & \esssup\widetilde h -\frac{\epsilon}{4} - \frac{\epsilon}{4} + \frac{\epsilon^2}{ 16\left|\esssup\widetilde h\right|}- \frac{\epsilon}{2} \\\geq & \esssup\widetilde h - \epsilon
\end{align*}
for all $N\geq \max\left\{N_1,\frac{1}{W}\right\}$. Noting that $\lambda_{l'}\left(\overline{\bm C}_N \right) \leq\lambda_{\rho(0)}\left( \overline{\bm C}_N\right) \leq \esssup\widetilde h$, we have
\begin{align*}
\left| \lambda_{\rho(0)}\left( \bm C_N\right) - \esssup\widetilde h \right|\leq \epsilon
\end{align*}
for all $N\geq N_0$. Since $\epsilon$ is arbitrary, we conclude
$$\lim_{N\rightarrow \infty}\lambda_{\rho(0)}\left( \overline{\bm C}_N\right) = \esssup\widetilde h.$$ With a similar argument, we have $$\lim_{N\rightarrow \infty}\lambda_{\rho{(N-1)}}\left( \overline{\bm C}_N\right) = \essinf\widetilde h.$$

Noting that $\lambda_{\rho(0)}\left( \overline{\bm C}_N\right)\leq \lambda_{0}\left( \bm H_N\right)\leq  \esssup\widetilde h$ and $\essinf \widetilde h\leq \lambda_{N-1}(\bm H_N) < \lambda_{\rho(N-1)}(\overline {\bm C}_N)$ (see Lemma~\ref{lem:bounds Pearl's eigenvalues}), we obtain
\begin{align*}&
\lim_{N\rightarrow \infty}\lambda_{\rho(0)}\left( \overline{\bm C}_N\right) = \lim_{N\rightarrow \infty}\lambda_0\left( \bm H_N\right) = \esssup\widetilde h\\
&\lim_{N\rightarrow \infty}\lambda_{\rho(N-1)}\left( \overline{\bm C}_N\right) = \lim_{N\rightarrow \infty}\lambda_{N-1}\left( \bm H_N\right) = \essinf\widetilde h.
\end{align*}
\cqfd

\section{Proof of Lemma \ref{lem:Pearl circulant for window function}}\label{prf:Pearl circulant for window function}
The following result indicates that the main lobe of the Dirichlet kernel contains most of its energy.
\begin{lem}
Let $D_N(f) = \frac{\sin\left(\pi N f  \right)}{\sin\left(\pi f \right)}$ be the Dirichlet kernel. Then
\begin{align*}
\int_{0}^{\frac{1}{N}} \left|D_N(f)\right|^2 df \geq 0.45N.
\end{align*}
\label{lem:main energy of Dirichlet kernel}\end{lem}
\vspace{0.3cm}
\noindent{\textbf{Proof}} (of Lemma \ref{lem:main energy of Dirichlet kernel} ). Noting that $\left|D_N(f)\right| = \frac{\left|\sin(\pi Nf) \right|}{\left| \sin(\pi f) \right|} \geq \frac{\left|\sin(\pi Nf) \right|}{\left| \pi f \right|}$, we have
\begin{align*}
\int_{0}^{\frac{1}{N}} \left|D_N(f)\right|^2 df &\geq \int_{0}^{\frac{1}{N}} \left|\frac{\sin(\pi Nf)}{\pi f}\right|^2 df = \frac{N}{\pi}\int_0^\pi \left|\frac{\sin(f)}{f}\right|^2 df\\
& = \frac{N}{\pi}\int_0^\pi \frac{\frac{1-\cos(2f)}{2}}{f^2}df \\
& = \frac{N}{2\pi} \int_0^{\pi} \sum_{k=1}^\infty \frac{(-1)^{k+1}(2f)^{2k}}{(2k)!f^2}df\\
& = \frac{N}{\pi}\sum_{k=1}^\infty \frac{ (-1)^{k+1}(2\pi)^{2k-1}}{(2k)!(2k-1)}\\
& \geq \frac{N}{\pi}\sum_{k=1}^{8}\frac{(-1)^{(k+1)}(2\pi)^{2k-1}}{\left(2k \right)!\left(2k-1 \right)} \\ & \geq 0.45N,
\end{align*}
where the third line follows from the common Taylor series $\cos(2f) = \sum_{k=0}^\infty (-1)^k\frac{(2f)^{2k}}{(2k)!}$, and the fifth line holds from the following inequality
\begin{align*}
&\sum_{k=9}^{\infty}\frac{(-1)^{(k+1)}(2\pi)^{2k-1}}{\left(2k \right)!\left(2k-1 \right)} \\
&=\sum_{n=5}^\infty \frac{(2\pi)^{4n-3}}{\left(4n-2 \right)!\left(4n-3 \right)} - \frac{(2\pi)^{4n-1}}{\left(4n \right)!\left(4n-1 \right)}\geq 0
\end{align*}
since
\begin{align*}
\frac{(2\pi)^{4n-1}}{\left(4n \right)!\left(4n-1 \right)} &\leq
\frac{(2\pi)^2}{(4n)(4n-1)}\frac{(2\pi)^{4n-3}}{\left(4n-2 \right)!\left(4n-3 \right)}\\
& \leq \frac{(2\pi)^{4n-3}}{\left(4n-2 \right)!\left(4n-3 \right)}
\end{align*}
for all $n\geq 2$.
\cqfd
Suppose $N$ is a multiple of $4$. Note that
\begin{align*}
\lambda_l(\overline{\bm C}_N) &= \int_{0}^{1}\left|\frac{1}{\sqrt{N}}\widetilde {\bm e}_{l/N}\right|^2\widetilde h(f)df \\ & = \int_0^{\frac{1}{4}} \left|\frac{1}{\sqrt{N}}\widetilde {\bm e}_{l/N}(f)\right|^2 df + \int_{\frac{3}{4}}^1\left|\frac{1}{\sqrt{N}}\widetilde {\bm e}_{l/N}(f)\right|^2 df,
\end{align*}
where $\widetilde e_{l/N}(f)$ is defined in~\eqref{eq:DTFT e}.
If $l = N/4$, $\left|\frac{1}{\sqrt{N}}\widetilde {\bm e}_{l/N}(f) \right| = \frac{1}{\sqrt N}\left| D_N(f - \frac{1}{4}) \right|$. Thus
\begin{align*}
&\lambda_{N/4}(\overline{\bm C}_N)\\ = & \frac{1}{N}\int_0^{\frac{1}{4}} \left|D_N(f-\frac{1}{4})\right|^2 df + \frac{1}{N}\int_{\frac{3}{4}}^1\left|D_N(f-\frac{1}{4})\right|^2 df \\= & \frac{1}{N}\int_0^{1/2}\left|D_N(f)\right|^2 df =\frac{1}{N}\frac{1}{2}\int_0^{1}\left|D_N(f)\right|^2 df = \frac{1}{2}.
\end{align*}
Similarly, we have $\lambda_{3N/4}(\overline{\bm C}_N) = \frac{1}{2}$.

Now for any $l\in[N], ~\frac{l}{N}\in [0,\frac{1}{4})\cup(\frac{3}{4},1]$, the main lobe of $D_N(f - \frac{l}{N})$ is inside the interval $[0,\frac{1}{4}]\cup[\frac{3}{4},1]$. Thus
\begin{align*}
&\lambda_{l}(\overline{\bm C}_N)\\
=& \frac{1}{N}\int_0^{\frac{1}{4}} \left|D_N(f-\frac{l}{N})\right|^2 df + \frac{1}{N}\int_{\frac{3}{4}}^1\left|D_N(f-\frac{l}{N})\right|^2 df \\ \geq & \frac{2}{N}\int_{0}^{\frac{1}{N}} \left|D_N(f)\right|^2 df\geq 0.9.
\end{align*}

Similarly, for any $l\in[N], ~\frac{l}{N}\in (\frac{1}{4},\frac{3}{4})$, we have
\begin{align*}
&\lambda_{l}(\overline{\bm C}_N)\\ = &\frac{1}{N}\int_0^{\frac{1}{4}} \left|D_N(f-\frac{l}{N})\right|^2 df + \frac{1}{N}\int_{\frac{3}{4}}^1\left|D_N(f-\frac{l}{N})\right|^2 df \\ \leq & 1 - \frac{2}{N}\int_{0}^{\frac{1}{N}} \left|D_N(f)\right|^2 df\leq 0.1.
\end{align*}

The proof is completed by noting that $0\leq\lambda_l(\overline{\bm C}_N)\leq 1$ for all $l\in[N]$. \cqfd

\bibliographystyle{ieeetr}
\bibliography{ToeplitzBibfile}

\begin{thebibliography}{10}

\bibitem{grenander1958toeplitz}
U.~Grenander and G.~Szeg\H{o}, {\em \uppercase{T}oeplitz Forms and Their
  Applications}, vol.~321.
\newblock Univ of California Press, 1958.

\bibitem{gray1972asymptotic}
R.~Gray, ``On the asymptotic eigenvalue distribution of \uppercase{T}oeplitz
  matrices,'' {\em IEEE Trans. Inf. Theory}, vol.~18, no.~6, pp.~725--730,
  1972.

\bibitem{pearl1973coding}
J.~Pearl, ``On coding and filtering stationary signals by discrete
  \uppercase{F}ourier transforms,'' {\em IEEE Trans. Inf. Theory}, vol.~19,
  no.~2, pp.~229--232, 1973.

\bibitem{makhoul1975linear}
J.~Makhoul, ``Linear prediction: A tutorial review,'' {\em Proc. IEEE},
  vol.~63, no.~4, pp.~561--580, 1975.

\bibitem{kailath1978inverses}
T.~Kailath, A.~Vieira, and M.~Morf, ``Inverses of \uppercase{T}oeplitz
  operators, innovations, and orthogonal polynomials,'' {\em SIAM Rev.},
  vol.~20, no.~1, pp.~106--119, 1978.

\bibitem{deift2012Spectrum}
P.~Deift, A.~Its, and I.~Krasovsky, ``Eigenvalues of \uppercase{T}oeplitz
  matrices in the bulk of the spectrum,'' {\em Bull. Inst. Math., Acad. Sin.
  (N.S.)}, pp.~437--461, 2012.

\bibitem{gray2005toeplitz}
R.~M. Gray, ``Toeplitz and circulant matrices: A review,'' {\em Commun. Inf.
  Theory}, vol.~2, no.~3, pp.~155--239, 2005.

\bibitem{avram1988bilinear}
F.~Avram, ``On bilinear forms in gaussian random variables and
  \uppercase{T}oeplitz matrices,'' {\em Probab. Theory Related Fields},
  vol.~79, no.~1, pp.~37--45, 1988.

\bibitem{parter1986distribution}
S.~V. Parter, ``On the distribution of the singular values of
  \uppercase{T}oeplitz matrices,'' {\em Linear Algebra Appl.}, vol.~80,
  pp.~115--130, 1986.

\bibitem{tyrtyshnikov1996unifying}
E.~E. Tyrtyshnikov, ``A unifying approach to some old and new theorems on
  distribution and clustering,'' {\em Linear Algebra Appl.}, vol.~232,
  pp.~1--43, 1996.

\bibitem{zamarashkin1997distribution}
N.~L. Zamarashkin and E.~E. Tyrtyshnikov, ``Distribution of eigenvalues and
  singular values of \uppercase{T}oeplitz matrices under weakened conditions on
  the generating function,'' {\em Sbornik: Mathematics}, vol.~188, no.~8,
  pp.~1191--1201, 1997.

\bibitem{sakrison1969extension}
D.~Sakrison, ``An extension of the theorem of \uppercase{K}ac,
  \uppercase{M}urdock and \uppercase{S}zeg{\H o} to \uppercase{$N$}
  dimensions,'' {\em IEEE Trans. Inf. Theory}, vol.~15, no.~5, pp.~608--610,
  1969.

\bibitem{gazzah2001asymptotic}
H.~Gazzah, P.~A. Regalia, and J.-P. Delmas, ``Asymptotic eigenvalue
  distribution of block \uppercase{T}oeplitz matrices and application to blind
  \uppercase{SIMO} channel identification,'' {\em IEEE Trans. Inf. Theory},
  vol.~47, no.~3, pp.~1243--1251, 2001.

\bibitem{gutierrez2008asympBolckToeplitz}
J.~Guti{\'e}rrez-Guti{\'e}rrez and P.~M. Crespo, ``Asymptotically equivalent
  sequences of matrices and hermitian block \uppercase{T}oeplitz matrices with
  continuous symbols: Applications to \uppercase{MIMO} systems,'' {\em IEEE
  Trans. Inf. Theory}, vol.~54, no.~12, pp.~5671--5680, 2008.

\bibitem{Bogoya2015maximum}
J.~Bogoya, A.~Böttcher, S.~Grudsky, and E.~Maximenko, ``Maximum norm versions
  of the \uppercase{S}zeg{\H o} and \uppercase{A}vram– \uppercase{P}arter
  theorems for \uppercase{T}oeplitz matrices,'' {\em J. Approx. Theory},
  vol.~196, no.~0, pp.~79 -- 100, 2015.

\bibitem{dembo1988bounds}
A.~Dembo, ``Bounds on the extreme eigenvalues of positive-definite
  \uppercase{T}oeplitz matrices,'' {\em IEEE Trans. Inf. Theory}, vol.~34,
  no.~2, pp.~352--355, 1988.

\bibitem{laudadio2008lowerBoundToeplitz}
T.~Laudadio, N.~Mastronardi, and M.~Van~Barel, ``Computing a lower bound of the
  smallest eigenvalue of a symmetric positive-definite \uppercase{T}oeplitz
  matrix,'' {\em IEEE Trans. Inf. Theory}, vol.~54, no.~10, pp.~4726--4731,
  2008.

\bibitem{mitola1999cognitive}
J.~Mitola~III and G.~Q. Maguire~Jr, ``Cognitive radio: \uppercase{M}aking
  software radios more personal,'' {\em IEEE Personal Commun.}, vol.~6, no.~4,
  pp.~13--18, 1999.

\bibitem{haykin2005cognitive}
S.~Haykin, ``Cognitive radio: \uppercase{B}rain-empowered wireless
  communications,'' {\em IEEE J. Select. Areas Commun.}, vol.~23, no.~2,
  pp.~201--220, 2005.

\bibitem{zeng2009eigenvalueSpectrumSensing}
Y.~Zeng and Y.-C. Liang, ``Eigenvalue-based spectrum sensing algorithms for
  cognitive radio,'' {\em IEEE Trans Commun.}, vol.~57, no.~6, pp.~1784--1793,
  2009.

\bibitem{strang1986proposal}
G.~Strang, ``A proposal for \uppercase{T}oeplitz matrix calculations,'' {\em
  Stud. Appl. Math.}, vol.~74, no.~2, pp.~171--176, 1986.

\bibitem{chan1988optimal}
T.~F. Chan, ``An optimal circulant preconditioner for \uppercase{T}oeplitz
  systems,'' {\em SIAM J. Sci. Statis. Comput.}, vol.~9, no.~4, pp.~766--771,
  1988.

\bibitem{korner1989fourier}
T.~W. K{\"o}rner, {\em Fourier analysis}.
\newblock Cambridge university press, 1989.

\bibitem{Horn1985MatrixAnalysis}
R.~A. Horn and C.~R. Johnson, eds., {\em Matrix Analysis}.
\newblock New York, NY, USA: Cambridge University Press, 1986.

\bibitem{trench1989numerical}
F.~W. Trench, ``Numerical solution of the eigenvalue problem for hermitian
  toeplitz matrices,'' {\em SIAM J. Matrix Anal. Appl.}, vol.~10, no.~2,
  pp.~135--146, 1989.

\bibitem{luk2000fast}
F.~T. Luk and S.~Qiao, ``A fast eigenvalue algorithm for hankel matrices,''
  {\em Linear Algebra Appl.}, vol.~316, no.~1-3, pp.~171--182, 2000.

\bibitem{chan1989circulant}
R.~H. Chan, ``Circulant preconditioners for hermitian \uppercase{T}oeplitz
  systems,'' {\em SIAM J. Matrix Anal. Appl.}, vol.~10, no.~4, pp.~542--550,
  1989.

\bibitem{chan1989toeplitz}
R.~H. Chan and G.~Strang, ``Toeplitz equations by conjugate gradients with
  circulant preconditioner,'' {\em SIAM J. Sci. Statis. Comput.}, vol.~10,
  no.~1, pp.~104--119, 1989.

\bibitem{chan1991spectra}
R.~H. Chan, X.-Q. Jin, and M.-C. Yeung, ``The spectra of super-optimal
  circulant preconditioned \uppercase{T}oeplitz systems,'' {\em SIAM J. Numer.
  Anal.}, vol.~28, no.~3, pp.~871--879, 1991.

\bibitem{trench2012elementary}
W.~F. Trench, ``An elementary view of \uppercase{W}eyl's theory of equal
  distribution,'' {\em Amer. Math. Monthly}, vol.~119, no.~10, pp.~852--861,
  2012.

\bibitem{zizler2002finer}
P.~Zizler, R.~A. Zuidwijk, K.~F. Taylor, and S.~Arimoto, ``A finer aspect of
  eigenvalue distribution of selfadjoint band \uppercase{T}oeplitz matrices,''
  {\em SIAM J. Matrix Anal. Appl.}, vol.~24, no.~1, pp.~59--67, 2002.

\bibitem{Slepian78DPSS}
D.~Slepian, ``Prolate \uppercase{S}pheroidal \uppercase{W}ave
  \uppercase{F}unctions, \uppercase{F}ourier analysis, and uncertainty.
  \uppercase{V}- \uppercase{T}he discrete case,'' {\em Bell Syst. Tech. J.},
  vol.~57, no.~5, pp.~1371--1430, 1978.

\bibitem{zhu2015approximating}
Z.~Zhu and M.~B. Wakin, ``Approximating sampled sinusoids and multiband signals
  using multiband modulated \uppercase{DPSS} dictionaries,'' {\em {\em to
  appear in} J. Fourier Anal. Appl.}, 2016.

\bibitem{van2000asymptotic}
A.~W. Van~der Vaart, {\em Asymptotic Statistics}, vol.~3.
\newblock Cambridge university press, 2000.

\bibitem{apostol1974mathematical}
T.~M. Apostol, {\em Mathematical Analysis (2nd ed.)}.
\newblock Addison Wesley Publishing Company, 1974.

\end{thebibliography}

\end{document}